\documentclass[times,twocolumn,final,sort&compress,numbers]{elsarticle}

\usepackage{cnf}

\usepackage{framed,multirow}

\usepackage{amssymb}
\usepackage{latexsym}

\usepackage{url}
\usepackage{xcolor}
\definecolor{newcolor}{rgb}{.8,.349,.1}

\usepackage{hyperref}

\usepackage[switch,pagewise]{lineno} 
\usepackage[fleqn]{amsmath}

\usepackage{multirow}  
\usepackage{array}   

\usepackage[version=4]{mhchem}

\journal{Preprint submitted to arXiv}

\begin{document}

\verso{D. Yang et al.}

\begin{frontmatter}

\title{Thermochemically Constrained Multiscale Laplace Neural Operator to Learn Stiff Chemical Kinetics with Conditional Diffusion Model Sampling}%

\author[1]{Dong \snm{Yang}}


\author[1]{Xiongbo \snm{Duan}}
    
\author[1]{Min \snm{Zhang}\corref{cor1}}
\cortext[cor1]{Corresponding authors at: School of Energy Science and Engineering, Central South University, Changsha, 410083, People's Republic of China.}
\emailauthor{minzhang@csu.edu.cn}{Min Zhang}

\author[1]{Yuqiang \snm{Li}}

\address[1]{School of Energy Science and Engineering, Central South University, Changsha, 410083, People's Republic of China}

\begin{abstract}
Stiff chemical kinetics is a major computational bottleneck in multidimensional reacting-flow simulations, 
while data-driven solvers remain sensitive to training-state coverage and the representation of multiscale chemical dynamics. 
This work combines conditional diffusion sampling with a thermochemically constrained multiscale Laplace neural operator (TC-MLNO) for chemistry ODE integration. 
The diffusion model learns the equivalence-ratio-conditioned joint distribution of temperature and species from low-dimensional flame data, 
augmenting the thermochemical state space without independently prescribed perturbations. TC-MLNO represents transient, 
non-periodic responses through pole--residue Laplace layers and assigns fast, medium, and slow species to separate branches. 
Thermochemical constraints are incorporated through training losses for mass conservation, elemental conservation, and heat release rate. 
The framework is evaluated in one-dimensional premixed flames, a two-dimensional temporally developing jet, and a three-dimensional turbulent burner flame. 
With equal-sized datasets, diffusion sampling reduces the $95^{th}$-percentile distance to the base flame manifold by up to 37.5\% and the 
a posteriori full-species macro-average error by up to 92.7\% compared with random sampling, with the largest benefits observed for minor species and radicals. 
TC-LNO represents modal attenuation more accurately than DeepONet baseline, 
yielding smaller errors for 84.9\% of the retained modes and reducing the $95^{th}$-percentile error for the most strongly attenuated modes from 0.372 to 0.081.
In multidimensional flames, TC-MLNO more accurately preserves reaction-zone locations, heat-release structures, and chemically sensitive species than DeepONet. 
The errors of representative radical and nitrogen-containing species in the three-dimensional flame decrease by 88.0\%--88.9\%. Relative to direct stiff integration using CVODE, 
TC-MLNO reduces the mean chemistry and total solution times by 90.1\% and 62.6\%, respectively. 
These results demonstrate that distribution-aware sampling and Laplace neural operator learning provide an accurate and efficient alternative for stiff chemical ODE 
integration in reacting-flow simulations.
\end{abstract}


\end{frontmatter}


\section*{Novelty and significance statement}
The novelty of this work lies in using a conditional diffusion model to improve thermochemical-state sampling and in developing 
a thermochemically constrained multiscale Laplace neural operator (TC-MLNO) framework 
for accurate prediction of stiff chemical kinetics. Unlike independently prescribed random perturbations, 
conditional diffusion sampling learns the equivalence-ratio-conditioned joint distribution of temperature and species, 
providing augmented states that better preserve flame-manifold correlations and minor species behavior. 
TC-MLNO combines pole–residue Laplace layers with reaction-timescale-specific branches 
while incorporating thermochemical constraints to represent non-periodic transient responses across fast, medium, and slow species.
This work is significant because the resulting framework improves chemical prediction in multidimensional reacting flows while substantially reducing the cost of 
repeated stiff ODE integration, providing an accurate and efficient approach to combustion-chemistry acceleration.

\section{Introduction}
\label{sec:Introduction}
Predictive design of modern combustion systems increasingly relies on high-fidelity computational fluid dynamics (CFD) with finite-rate chemistry. 
However, detailed mechanisms give rise to high-dimensional, strongly coupled, and stiff ordinary differential equations (ODEs) whose repeated integration at every grid point and time step constitutes a major computational bottleneck \cite{luAccommodatingRealisticFuel2009,ramanEmergingTrendsNumerical2019}. 
Chemical kinetics can account for more than 80\% of the total cost in simulations involving tens of species \cite{liComprehensiveDeepLearning2025}. 
Neural networks provide an alternative by replacing online chemical integration with inexpensive learned evaluations, as demonstrated from early multilayer perceptrons to CFD-coupled chemistry surrogates \cite{christoArtificialNeuralNetwork1996,gcerri.KineticCombustionNeural2003,ihmeCombustionMachineLearning2022}. 
Their reliability, however, depends critically on both the quality of the training data and the representation of transient chemical evolution.

Training data for neural chemistry models are commonly generated by sampling initial temperature, pressure, and composition, and then collecting thermochemical states from homogeneous reactors, canonical laminar flames, or reacting-flow simulations \cite{anExploreArtificialNeural2022,chiOntheflyArtificialNeural2021}. 
Trajectory- and manifold-based samples preserve physically realizable correlations but are confined to the configurations from which they are obtained and may generalize poorly to thermochemical states that deviate from the sampled trajectories or manifolds because of transport, mixing, and accumulated prediction errors. 
Conversely, global random sampling broadens the explored state space but inefficiently represents the multiscale distributions of minor species and radicals. 
Zhang et al. \cite{zhangMultiscaleSamplingMethod2022} systematically compared different sampling strategies and showed that manifold data may lose robustness under perturbations, whereas broader random or generative samples can still miss small-scale intermediate species. 
To improve robustness to deviations from the sampled manifold, Ding et al. \cite{dingMachineLearningTabulation2021} combined flamelet and random data, Han et al. \cite{hanImprovedApproachMore2022} perturbed simulated trajectories, and Li et al. \cite{liComprehensiveDeepLearning2025} augmented canonical-flame base states for multiple fuels. 
Nevertheless, these methods require empirically selected perturbation amplitudes and repeated filtering or normalization. Small perturbations add mostly redundant samples, whereas large or independently imposed variations can distort correlations among temperature and species and yield unrealistic thermochemical states. 
Generative models offer a distribution-learning alternative. 
In particular, diffusion models approximate complex joint distributions through a forward noising process and a learned reverse denoising process \cite{yangDiffusionModelsComprehensive2024}, and have demonstrated the ability to synthesize correlated multivariate data while retaining dependencies among variables \cite{villaizan-valleladoDiffusionModelsTabular2025}. 
This provides a systematic route to augment thermochemical samples according to learned cross-variable dependencies rather than prescribed local perturbations, although conservation and physical admissibility must still be enforced.

In parallel, the surrogate architecture determines how chemical dynamics are represented. 
Conventional networks approximate reaction rates or discrete state mappings, whereas physics-informed and neural ODE formulations incorporate governing laws or continuous dynamics. 
Ji et al. \cite{jiStiffPINNPhysicsinformedNeural2021} examined physics-informed networks for stiff chemical kinetics, Zhang et al. \cite{zhangCRKPINNPhysicsinformedNeural2024} imposed kinetic and conservation relations during training, and Saito et al. \cite{saitoDatadrivenSimulationAmmonia2023} applied neural ODE to ammonia combustion. 
These approaches can improve physical consistency or temporal continuity, although stiffness and limited training-state coverage remain challenges for data-driven chemistry surrogates. 
Neural operators offer a complementary formulation by learning mappings between input and solution functions, thereby representing chemical evolution across initial conditions and temporal coordinates \cite{azizzadenesheliNeuralOperatorsAccelerating2024,wengExtendedFourierNeural2025}. 
Goswami et al. \cite{goswamiLearningStiffChemical2024a} used extended DeepONet to learn solution propagators for stiff benchmark systems, syngas kinetics, and a reacting jet flame. 
Kumar and Echekki \cite{kumarCombustionChemistryAcceleration2024} further developed a DeepONet-based chemistry scheme with flexible time increments and demonstrated it for hydrogen and n-dodecane kinetics. 
These studies establish the potential of operator learning for chemical ODEs, while also motivating representations better suited to transient dynamics. 
The Laplace neural operator (LNO) introduces a physically interpretable pole--residue relation among the input, kernel, and system response, representing transient and steady responses under zero and non-zero initial conditions \cite{caoLaplaceNeuralOperator2024}. 
These properties align with the transient, multiscale, and non-periodic evolution of combustion chemistry. 
This structure offers potential advantages for representing combustion chemistry. 
Further development of LNO is therefore needed to fully exploit its potential for learning combustion dynamics and chemical evolution.

To improve the quality and physical representativeness of training data while providing a neural operator suited to the transient, multiscale, 
and non-periodic dynamics of stiff chemistry, the present study has two aims: 
(i) to adopt a conditional diffusion sampling strategy that augments low-dimensional flame data while preserving the thermochemical correlations among temperature and species, 
and to quantify its advantages over random sampling; and (ii) to construct an thermochemically constrained multiscale Laplace neural operator (TC-MLNO) framework that combines pole--residue Laplace layers, 
reaction-timescale grouping, and physics-based loss terms, and to assess its accuracy and efficiency in one-dimensional and multidimensional reacting flows.

The remainder of this paper is structured as follows. Section~\ref{sec:methodology} formulates the chemical solution operator and describes the conditional diffusion sampling method, 
the TC-MLNO architecture, and the baseline model. Section~\ref{sec:ResultsDiscussion} evaluates the sampled-state distributions and chemistry prediction models through modal 
analysis and one-dimensional, two-dimensional, and three-dimensional reacting-flow simulations, together with computational-cost measurements. 
Section~\ref{sec:Conclusions} summarizes the principal findings and provides an outlook for future work.

\section{Methodology}
\label{sec:methodology}

\begin{figure*}[!t]
\centering
\includegraphics[scale=0.45]{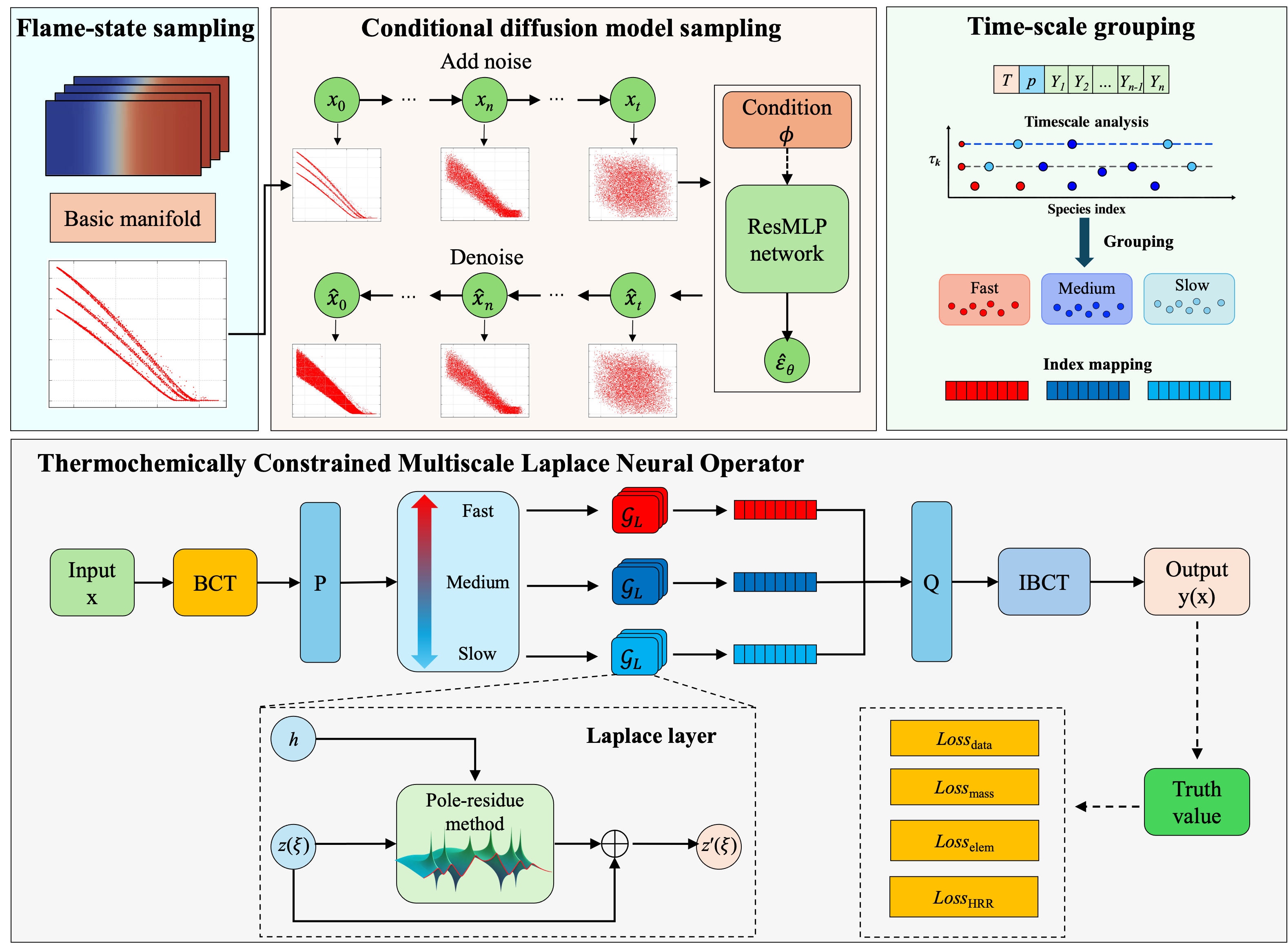}
\caption{The framework map of the conditional diffusion model sampling and thermochemically constrained multiscale Laplace neural operator model.\label{fig:framework}}
\end{figure*}

\subsection{Stiff chemical kinetics systems}
\label{subsec:kinetics_propagator}

For a homogeneous reacting system, the thermochemical state at time level $n$ is
\begin{equation}
    \mathbf{q}^{n}
    =
    \left[T^{n},p^{n},\mathbf{Y}^{n}\right]^{\mathrm T},
    \qquad
    \mathbf{Y}^{n}
    =
    \left[Y^{n}_{1},\ldots,Y^{n}_{N_s}\right]^{\mathrm T},
    \label{eq:thermochemical_state}
\end{equation}
where $T$, $p$, and $Y_k$ are temperature, pressure, and species mass fraction, respectively, and $N_s$ is the number of species. Species evolve according to
\begin{equation}
    \frac{\mathrm{d}Y_k}{\mathrm{d}t}
    =
    \frac{W_k\dot{\omega}_k(\mathbf{q})}{\rho},
    \qquad k=1,\ldots,N_s,
    \label{eq:species_kinetics}
\end{equation}
where $W_k$, $\dot{\omega}_k$, and $\rho$ are the molecular weight, net molar production rate, and mixture density. The wide range of species time scales makes Eq.~\eqref{eq:species_kinetics} stiff and generally requires adaptive implicit integration with small internal steps.

Over a prescribed chemical time increment $\Delta t$, Eq.~\eqref{eq:species_kinetics} defines the finite-time solution operator
\begin{equation}
    \mathbf{Y}^{n+1}
    =
    \mathcal{S}_{\Delta t}\left(\mathbf{q}^{n}\right),
    \qquad
    \widehat{\mathbf{Y}}^{\,n+1}
    =
    \widehat{\mathcal{S}}_{\vartheta,\Delta t}
    \left(\mathbf{q}^{n}\right),
    \label{eq:solution_propagator}
\end{equation}
where $\widehat{\mathcal{S}}_{\vartheta,\Delta t}$ approximates the numerical operator $\mathcal{S}_{\Delta t}$ to avoid stiff integration during deployment.

\subsection{Conditional diffusion model sampling}
\label{subsec:conditional_sampling}

As a conventional baseline, random sampling augments flame states \cite{jiangInterpretableFrameworkDatadriven2021,readshawModelingTurbulentFlames2021} according to

\begin{align}
    T' & = T
    +{\alpha}\left(T_{\max}-T_{\min}\right)\xi,
    \label{eq:random_temperature}\\
    Y_I' & = Y_I
    +{\alpha}
    \left(Y_{I,\max}-Y_{I,\min}\right)\xi,
    \label{eq:random_inert}\\
    Y_k' & = Y_k^{\,1+\beta\xi},
    \qquad k\neq I,
    \label{eq:random_species}
\end{align}
where $I$ is the inert species, $\alpha=0.05$, $\beta=0.10$, and $\xi\in[-1,1]$ is random. This amplitude pair shows the best prediction performance among all the pairs examined by Zhang et al.~\cite{zhangPerformanceComparisonMachine2026}. 
Because temperature and species are perturbed with independently prescribed amplitudes, this approach can weaken their joint correlations and generate off-manifold states despite 
normalization and filtering. 
This work instead collects base states from low-dimensional flames at discrete equivalence ratios and trains a conditional diffusion model to generate correlated thermochemical states. 
The sampling procedure is illustrated in the sampling component of Fig.~\ref{fig:framework}.
These base states retain the coupled variations of temperature and species along the flame manifold, but their direct use restricts the training distribution to the sampled conditions. 
A conditional diffusion model is therefore trained to approximate the thermochemical-state distribution conditioned on the equivalence ratio $\phi$. 
After training, a prescribed condition $\phi^{\ast}$ can be supplied to generate a large number of thermochemical states within the conditional range represented by the base data.

Define $\mathbf{x}_0$ as a thermochemical state in the model space and define $r=1,\ldots,R$ as the diffusion step. 
The forward process progressively adds Gaussian noise according to \cite{yangDiffusionModelsComprehensive2024}
\begin{equation}
    \mathbf{x}_r
    =
    \sqrt{\bar{\alpha}_r}\,\mathbf{x}_0
    +
    \sqrt{1-\bar{\alpha}_r}\,\boldsymbol{\epsilon},
    \qquad
    \boldsymbol{\epsilon}\sim\mathcal{N}(\mathbf{0},\mathbf{I}),
    \label{eq:diffusion_forward}
\end{equation}
where
\begin{equation}
    \alpha_r=1-\beta_r,
    \qquad
    \bar{\alpha}_r=\prod_{\nu=1}^{r}\alpha_{\nu},
    \label{eq:diffusion_schedule}
\end{equation}
and $\beta_r$ is the prescribed noise schedule. A conditional residual multilayer perceptron (ResMLP) receives the noisy state $\mathbf{x}_r$, 
the diffusion step $r$, and the equivalence ratio $\phi$, and predicts the added noise as
\begin{equation}
    \widehat{\boldsymbol{\epsilon}}_r
    =
    \boldsymbol{\epsilon}_{\theta_{\mathrm D}}
    \left(\mathbf{x}_r,r,\phi\right).
    \label{eq:conditional_noise_prediction}
\end{equation}

Starting from Gaussian noise, the sampled state is reconstructed through the reverse process
\begin{equation}
    \mathbf{x}_{r-1}
    =
    \frac{1}{\sqrt{\alpha_r}}
    \left[
        \mathbf{x}_r
        -
        \frac{1-\alpha_r}{\sqrt{1-\bar{\alpha}_r}}
        \boldsymbol{\epsilon}_{\theta_{\mathrm D}}
        \left(\mathbf{x}_r,r,\phi\right)
    \right]
    +
    \sqrt{\beta_r}\,\boldsymbol{\eta}_r,
    \label{eq:diffusion_reverse}
\end{equation}
where $\theta_{\mathrm D}$ denotes the denoising-network parameters, $\boldsymbol{\eta}_r\sim\mathcal{N}(\mathbf{0},\mathbf{I})$ for $r>1$, and $\boldsymbol{\eta}_1=\mathbf{0}$. The equivalence ratio and diffusion-step embedding enter each ResMLP block, conditioning reconstruction on the specified flame condition. Each accepted random or diffusion sample is then advanced over one chemical time increment by a high-accuracy stiff integrator to obtain its training target.

\subsection{Thermochemically constrained multiscale Laplace neural operator (TC-MLNO)}
\label{subsec:timescale_lno}

\subsubsection{Box--Cox transformation}
\label{subsubsec:bct}

Because species mass fractions span several orders of magnitude, a Box--Cox transformation (BCT) is first applied \cite{boxAnalysisTransformations1964}:
\begin{equation}
    b_k
    =
    \mathcal{B}_{\lambda}(Y_k)
    =
    \begin{cases}
        \dfrac{Y_k^{\lambda}-1}{\lambda}, & \lambda\neq 0,\\[4pt]
        \log Y_k, & \lambda=0,
    \end{cases}
    \label{eq:bct_increment}
\end{equation}
where $\lambda=0.1$ in this work. The transformed state is standardized as
\begin{equation}
    \widetilde{\mathbf{q}}^{\,n}
    =
    \mathcal{N}
    \left(
        T^{n},p^{n},\mathbf{b}^{n}
    \right),
    \label{eq:lno_input}
\end{equation}
where $\mathcal{N}$ denotes componentwise standardization, $\Delta b_k=b_k^{n+1}-b_k^n$, and the standardized target is $\Delta\mathbf{b}^{\,\mathrm{std}}$.

\subsubsection{Multiscale reaction-timescale grouping}
\label{subsubsec:timescale_grouping}

The species are divided according to their characteristic destruction times \cite{zhangLearningTransientEvolution2024,zhengRFNOPINNPhysicsinformedNeural2026}. 
For species $k$, the reaction time scale is defined as
\begin{equation}
    \tau_k
    =
    \frac{C_k}{\dot{\omega}_k^{-}},
    \label{eq:reaction_timescale}
\end{equation}
where $C_k$ and $\dot{\omega}_k^{-}$ are its molar concentration and destruction rate, respectively. 
The time scales are evaluated over representative training states, and the species are ordered using their median values relative to the network time increment. 
Splits at the largest admissible gaps yield the fast, medium, and slow groups shown in Fig.~\ref{fig:framework}. 
Temperature and pressure are retained in the shared input state rather than assigned to a reaction-scale group, so that
every group is conditioned on the same thermochemical environment.

\subsubsection{TC-MLNO architecture}
\label{subsubsec:mlno_architecture}

The Laplace layers in TC-MLNO build on the LNO formulation of Cao et al. \cite{caoLaplaceNeuralOperator2024}. 
Let $v(\xi)$, $u(\xi)$, and $\kappa_{\theta_{\mathrm L}}(\xi)$ denote the input, output, and kernel of a generic operator, 
and denote their Laplace transforms by $V(s)=\mathcal{L}\{v(\xi)\}$, $U(s)=\mathcal{L}\{u(\xi)\}$, 
and $K_{\theta_{\mathrm L}}(s)=\mathcal{L}\{\kappa_{\theta_{\mathrm L}}(\xi)\}$, respectively. 
The Laplace-domain input--output relation is
\begin{equation}
    U(s)=K_{\theta_{\mathrm L}}(s)V(s),
    \label{eq:laplace_domain_relation}
\end{equation}
here, $s=\sigma+\mathrm{i}\omega$ is the complex Laplace variable and $\theta_{\mathrm L}$ denotes the parameters of the Laplace layers. 
Each Laplace layer parameterizes the transfer function in pole--residue form,
\begin{equation}
    K_{\theta_{\mathrm L}}(s)
    =
    \sum_{m=1}^{M}
    \frac{\gamma_m}{s-\mu_m},
    \label{eq:pole_residue_kernel}
\end{equation}
where $\mu_m$ and $\gamma_m$ are trainable poles and residues, respectively, and $M$ is the number of pole--residue modes.

In the present network, the standardized thermochemical state is first encoded into a shared latent vector:
\begin{equation}
    \mathbf{h}=\mathcal{P}_{\theta_{\mathrm L}}(\widetilde{\mathbf{q}}^{\,n}),
    \label{eq:latent_state_encoding}
\end{equation}
based on $\mathbf{h}$, an initial-state-dependent lifting map constructs the latent function:
\begin{equation}
    \mathbf{z}_{g,0}(\xi)=
    \mathcal{T}_{\theta_{\mathrm L},g}(\mathbf{h},\xi),
    \label{eq:latent_time_lifting}
\end{equation}
where $\xi\in[0,1]$ and $g\in\{\mathrm{fast},\mathrm{medium},\mathrm{slow}\}$. For the $\ell$th layer, $c=1,\ldots,d_z$ indexes latent channels of the layer-normalized function $\bar{\mathbf z}_{\ell}$. Channel-dependent poles and residues are used in $K_{\ell,c}(s)$. With $N_\xi$ local-coordinate points, the discrete coefficients $\alpha_{\ell,c,j}=N_\xi^{-1}\mathcal F_\xi\{\bar z_{\ell,c}\}_j$ give
\begin{equation}
    Z_{\ell,c}(s)
    =
    \sum_{j=0}^{N_{\xi}-1}
    \frac{\alpha_{\ell,c,j}}
    {s-\mathrm{i}\omega_j}.
    \label{eq:latent_laplace_expansion}
\end{equation}

Evaluating $K_{\ell,c}$ at $s_j=\mathrm{i}\omega_j$ and applying the inverse transform yields
\begin{equation}
    z'_{\ell,c}(\xi)
    =
    \operatorname{Re}
    \left\{
        \mathcal{F}_{\xi}^{-1}
        \left[
            K_{\ell,c}(\mathrm{i}\boldsymbol{\omega})
            \odot
            \mathcal{F}_{\xi}\{\bar{z}_{\ell,c}\}
        \right]
    \right\},
    \label{eq:discrete_laplace_operator}
\end{equation}
where $\boldsymbol{\omega}=[\omega_0,\ldots,\omega_{N_\xi-1}]^{\mathrm T}$, $\odot$ is elementwise multiplication, and $\operatorname{Re}$ retains the real part. 
In each independent timescale branch, this response, a local $1\times1$ linear map, and the projected initial state form a residual update. 
Three conditional Laplace layers are stacked. The prediction grouped by reaction timescale is
\begin{equation}
\begin{split}
    \widehat{\Delta\mathbf{b}}^{\,\mathrm{std}}
    &=
    \mathcal{R}
    \Bigl(
        \mathcal{Q}_{\mathrm{fast}}\mathcal{G}_{L,\mathrm{fast}}(\mathbf{h})
        \oplus
        \mathcal{Q}_{\mathrm{medium}}\mathcal{G}_{L,\mathrm{medium}}(\mathbf{h}) \\
    &\quad \oplus
        \mathcal{Q}_{\mathrm{slow}}\mathcal{G}_{L,\mathrm{slow}}(\mathbf{h})
    \Bigr),
\end{split}
\label{eq:grouped_lno_output}
\end{equation}
where $\mathcal G_{L,g}$ and $\mathcal Q_g$ are the stacked-layer map and the output projection for group $g$, 
$\oplus$ denotes concatenation, and $\mathcal R$ restores species order. The branch outputs are anchored to zero increment at $\xi=0$, 
and the endpoint at $\xi=1$ is used as the single-step prediction.

The thermochemical constraints in TC-MLNO are imposed through loss penalties during training. The training objective is described by four $L$ components: the data loss,
\begin{equation}
    L_{\mathrm{Data}}
    =\left\lVert
        \widehat{\Delta\mathbf{b}}^{\,\mathrm{std}}-\Delta\mathbf{b}^{\,\mathrm{std}}
      \right\rVert_1,
    \label{eq:lno_data_loss}
\end{equation}
the mass-conservation loss,
\begin{equation}
    L_{\mathrm{Mass}}
    =\left\lVert
        \sum_{k=1}^{N_p}\widehat{Y}_{k}^{\,n+1}-\sum_{k=1}^{N_p}Y_{k}^{n}
      \right\rVert_1,
    \label{eq:lno_mass_loss}
\end{equation}
the elemental-conservation loss,
\begin{equation}
    L_{\mathrm{Elements}}
    =\left\lVert
        \mathbf{A}\left(\widehat{\mathbf{Y}}^{\,n+1}-\mathbf{Y}^{n}\right)
      \right\rVert_1,
    \label{eq:lno_element_loss}
\end{equation}
and the heat-release-rate loss,
\begin{equation}
    L_{\mathrm{HRR}}
    =\left\lVert
        \frac{\widehat{\dot{Q}}-\dot{Q}}{Q_{\mathrm{scale}}}
      \right\rVert_1.
    \label{eq:lno_hrr_loss}
\end{equation}
Here, $\lVert\cdot\rVert_1$ is the mean absolute error, $N_p$ is the number of explicitly predicted species, $\mathbf A$ is the elemental mass-contribution matrix, and $Q_{\mathrm{scale}}$ scales the heat-release rate.
The loss ablation study demonstrates that incorporating these physical constraints improves the predictive performance of TC-MLNO, as shown in Supplementary A.

\subsubsection{Baseline DeepONet}
\label{subsubsec:deeponet_baseline}

\begin{figure*}[!t]
    \centering
    \includegraphics[width=0.9\textwidth]{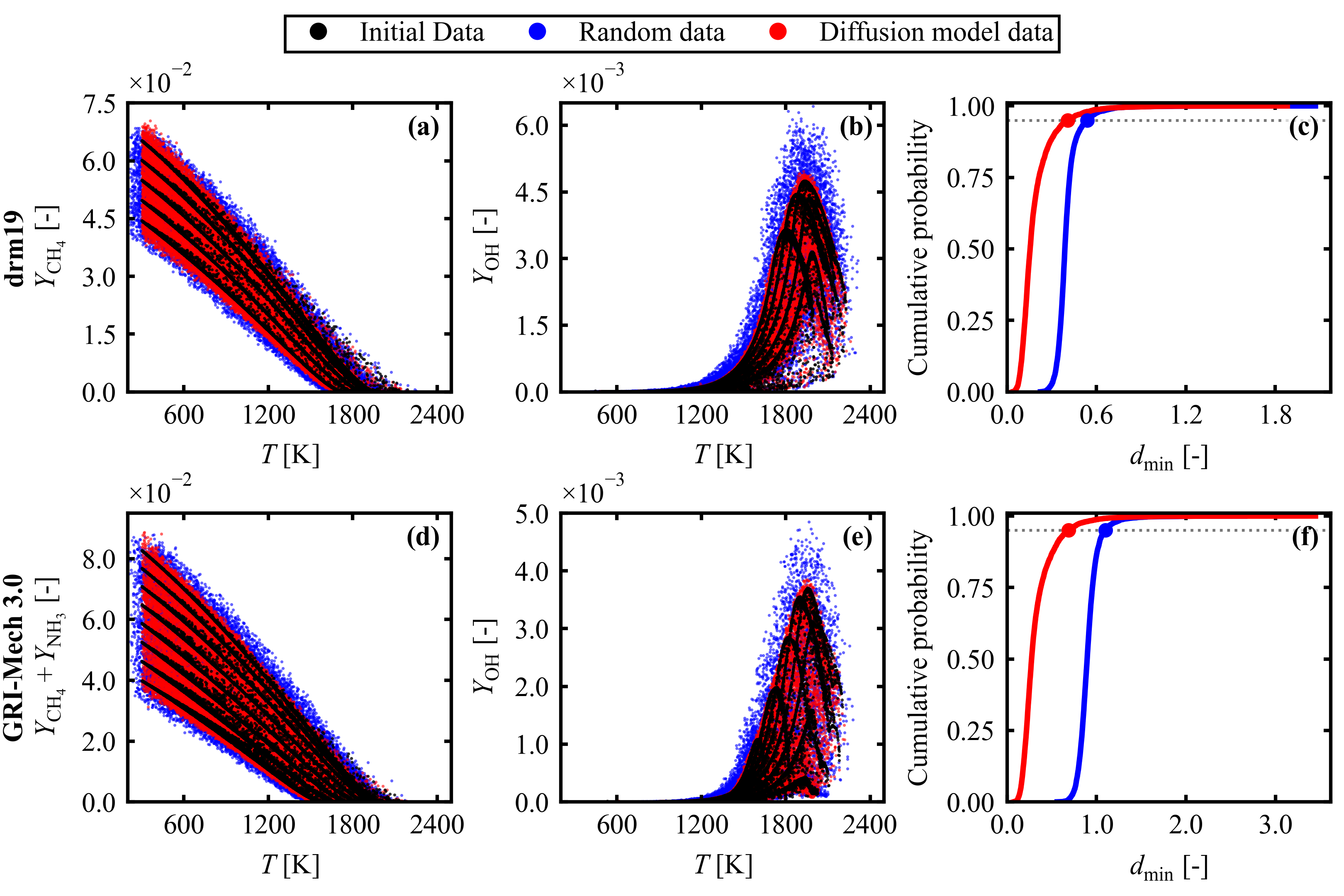}
    \caption{Comparison of the thermochemical states obtained using random sampling and conditional diffusion model sampling. 
    Panels (a) and (d) show the fuel mass fraction as a function of temperature; 
    panels (b) and (e) show $Y_{\mathrm{OH}}$ as a function of temperature; 
    and panels (c) and (f) show the cumulative probability distribution of the nearest-neighbor distance $d_{\min}$. 
    The horizontal dotted line and colored markers identify the $95^{th}$-percentile distance $D_{95}$.}
    \label{fig:sampling_state_coverage}
\end{figure*}

DeepONet serves as the operator-learning baseline following Refs.~\cite{wengExtendedFourierNeural2025,goswamiLearningStiffChemical2024a}. Its branch and trunk networks encode $\widetilde{\mathbf q}^{\,n}$ and $\xi$, respectively, and their inner product gives the standardized increment for species $k$:
\begin{equation}
    \widehat{\Delta b}_{k}^{\,\mathrm{std}}(\xi)
    =
    \sum_{j=1}^{m}
    \operatorname{branch}_{k,j}\!\left(\widetilde{\mathbf{q}}^{\,n}\right)
    \operatorname{trunk}_{j}(\xi),
\label{eq:deeponet_operator}
\end{equation}
where $m$ is the common feature dimension. Outputs are organized into species and local-time blocks before summation. Both networks use hyperbolic-tangent activation, with layer normalization only in the branch network and no output activation. All models are trained offline using stiff-integrator targets and evaluated through offline propagation and coupled reacting-flow calculations.

\section{Results and discussion}
\label{sec:ResultsDiscussion}

\subsection{Performance of the conditional diffusion model}
\label{subsec:sampling_performance}

\begin{figure}[!t]
    \centering
    \includegraphics[width=\columnwidth]{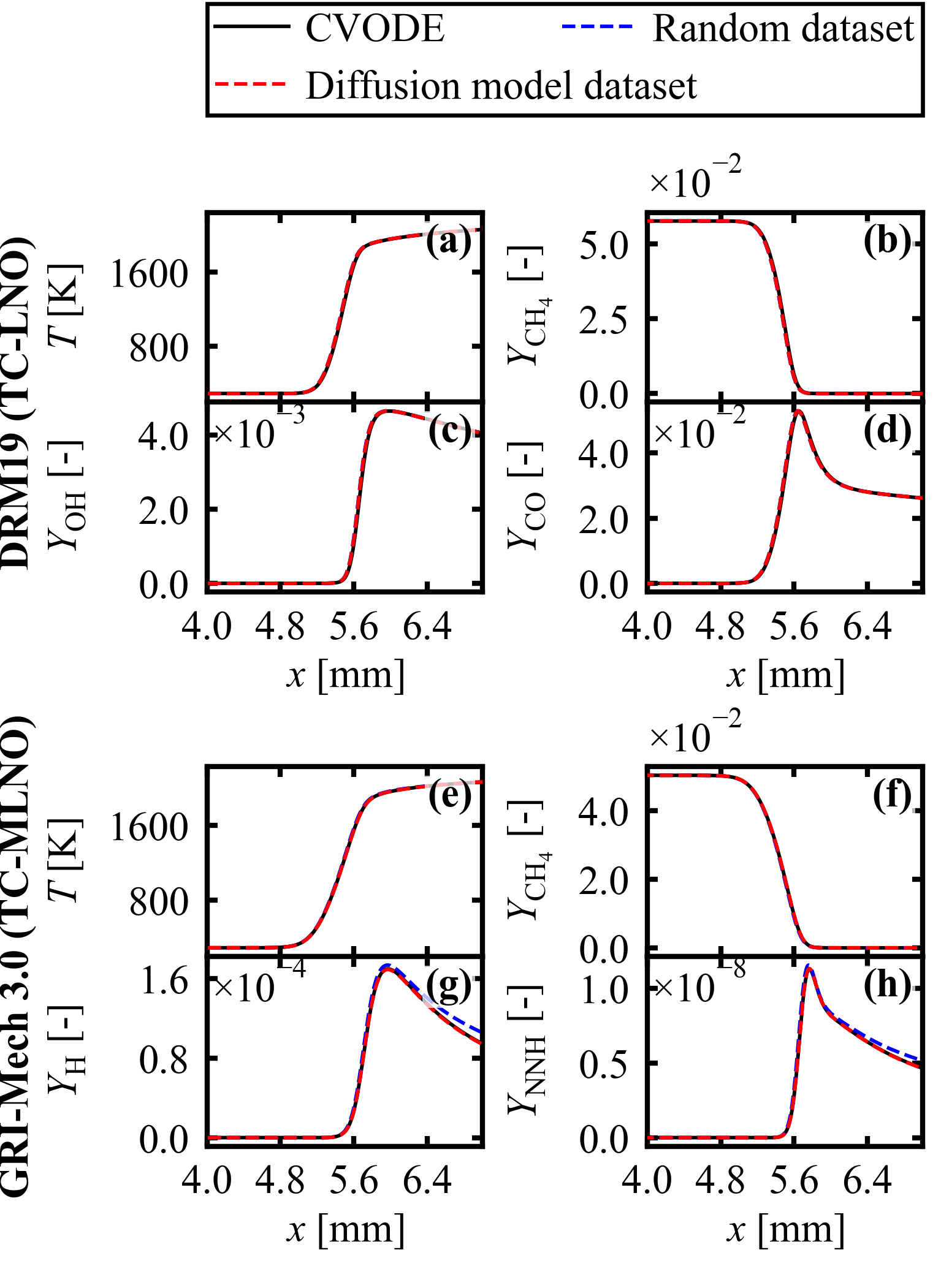}
    \caption{One-dimensional flame profiles obtained with CVODE and TC-LNO (DRM19) and TC-MLNO (GRI-Mech 3.0) trained using random and conditional diffusion datasets.}
    \label{fig:sampling_flame_profiles}
\end{figure}

The generated thermochemical states are evaluated using DRM19 \cite{kazakovDRM191995} (20 species, 82 reactions) and GRI-Mech~3.0 \cite{smithGRIMech301999} (53 species, 325 reactions). 
For DRM19, five one-dimensional \ce{CH4} flames with interval of 0.1 over $\phi=0.80$--$1.20$ form the base dataset, which is used to train the conditional diffusion model. 
The trained model then generates states at 200 equivalence-ratio conditions over $\phi=0.75$--$1.25$. 
For GRI-Mech~3.0, eight \ce{CH4}/\ce{NH3} ($3:1$) flames with interval of 0.1 over $\phi=0.60$--$1.30$ are used to train the conditional diffusion model, 
which subsequently generates states at 200 conditions over $\phi=0.65$--$1.35$ (Fig.~\ref{fig:sampling_state_coverage}). 
Sampling quality is assessed by comparing the generated states with the base flame states and by a posteriori flame calculations using equal-sized datasets obtained from 
random and conditional diffusion sampling. For the a posteriori flame calculations, 
the ungrouped thermochemically constrained Laplace neural operator (TC-LNO) and reaction-timescale-grouped TC-MLNO are adopted for DRM19 and GRI-Mech~3.0, respectively. 
The larger GRI-Mech~3.0 mechanism involves higher-dimensional kinetics and a wider separation of species time scales, 
for which the reaction-timescale-grouped branches of TC-MLNO are expected to provide improved predictive accuracy.

Figure~\ref{fig:sampling_state_coverage} compares the sampled distributions in representative temperature--species projections. 
Both sampling approaches expand the base flame states. 
However, the random samples form a relatively diffuse cloud around the initial flame data, with particularly broad deviations near the OH-rich reaction zone. 
By comparison, the diffusion samples remain more concentrated along the coupled variation of fuel consumption, temperature rise, and OH formation. 
This difference is quantified using the nearest-neighbor distance $d_{\min}$ from each initial one-dimensional flame state to the complete sampled database in the BCT 
and componentwise standardized state space \cite{NEURIPS2019_0234c510}. A smaller distance therefore indicates closer representation of the base flame manifold. 
For DRM19, the $95^{th}$-percentile distance $D_{95}$ decreases from 0.54 for random sampling to 0.41 for diffusion sampling, corresponding to a reduction of 24.0\%. 
For GRI-Mech~3.0, $D_{95}$ decreases from 1.10 to 0.69, a larger reduction of 37.5\%. 
The leftward shift of the cumulative probability distribution in both mechanisms demonstrates that the improvement is not restricted to a small subset of states. 
$D_{95}$ characterizes the representation of the base flame manifold, while the resulting prediction accuracy is evaluated next through a posteriori flame calculations.

The effect of this distributional difference on the coupled flame solution is shown in Fig.~\ref{fig:sampling_flame_profiles}. 
All models are trained using one million sampled states and evaluated at $\phi=1.05$, 
an unseen equivalence ratio that is not included among the base-state sampling conditions, with the CVODE solver providing the reference solution.
For DRM19, the TC-LNO predictions with random sampling and diffusion-model sampling are nearly indistinguishable from CVODE for temperature, fuel consumption, 
and the formation and decay of the representative intermediate and radical species. 
Thus, both sampling methods provide sufficient information for the TC-LNO to reproduce the main structure of this comparatively compact mechanism. 
A clearer separation is observed for GRI-Mech~3.0. Although both TC-MLNO models capture the flame position and the consumption of \ce{CH4}, 
the model trained with random samples develops visible deviations in the post-flame tails of H and NNH. 
The TC-MLNO predictions with diffusion-model sampling remain closer to CVODE through their peaks and decay regions. 
The principal benefit of diffusion sampling in this case is therefore not a substantial displacement of the overall flame front, 
but a more accurate reconstruction of minor species and long-tailed post-flame behavior in the higher-dimensional mechanism.

To quantify the overall a posteriori accuracy beyond the quantities selected for Fig.~\ref{fig:sampling_flame_profiles}, 
the full-species macro-average relative spatial error is defined as
\begin{equation}
    E_{\mathrm{macro}}
    =
    \frac{100.0\%}{N_s}
    \sum_{k=1}^{N_s}
    \frac{\left\|Y_k^{\mathrm{ML}}-Y_k^{\mathrm{CVODE}}\right\|_2}
    {\left\|Y_k^{\mathrm{CVODE}}\right\|_2}.
    \label{eq:sampling_macro_l2}
\end{equation}

As summarized in Table~\ref{tab:sampling_coverage_accuracy}, diffusion sampling reduces the DRM19 macro-average species error from 5.3\% to 4.1\%, or by 21.7\%. 
The improvement is substantially larger for GRI-Mech~3.0, for which the error decreases from 5.5\% to 0.4\%, representing a 92.7\% reduction. 
The simultaneous reductions in $D_{95}$ and $E_{\mathrm{macro}}$ for both mechanisms connect the 
closer representation of the flame-state distribution with improved coupled prediction. 
The species-resolved comparison in Fig~SB.~1 of the Supplementary B also shows that conditional diffusion sampling reduces errors in temperature and all displayed species for GRI-Mech~3.0.
The much stronger response for GRI-Mech~3.0 also indicates that sampling quality becomes increasingly important as the dimension and multiscale character of the chemical state increase.

\begin{table}[!t]
    \centering
    \caption{Sampling-state coverage and a posteriori full-species macro-average relative $L_2$ error for the steady one-dimensional flames at $\phi=1.05$.}
    \label{tab:sampling_coverage_accuracy}
    \fontsize{8pt}{9pt}\selectfont
    \setlength{\tabcolsep}{4pt}
    \begin{tabular}{l l c c}
        \hline
        Mechanism & Dataset & $D_{95}$ & $E_{\mathrm{macro}}$ (\%) \\
        \hline
        \multirow{2}{*}{DRM19}
        & Random          & 0.54 & 5.3 \\
        & Diffusion model & 0.41 & 4.1 \\
        \hline
        \multirow{2}{*}{GRI~3.0}
        & Random          & 1.10 & 5.5 \\
        & Diffusion model & 0.69 & 0.4 \\
        \hline
    \end{tabular}
\end{table}

The above results show that broadening the sampled range alone is insufficient to characterize sampling quality. 
Random perturbations generate admissible off-manifold states, but independently prescribed perturbations can weaken the correlations among temperature and species. 
The conditional diffusion model instead learns their joint distribution and produces an augmented database that remains more consistent 
with the reacting-state manifold. Having established the effectiveness of the training data, 
the following section examines the accuracy of the Laplace neural operator in predicting chemical kinetics.

\subsection{Performance of the TC-LNO and TC-MLNO chemistry prediction networks}
\label{subsec:lno_mlno_performance}

\subsubsection{TC-LNO compared with DeepONet}
\label{subsubsec:lno_deeponet_results}

\begin{figure*}[!t]
    \centering
    \includegraphics[width=0.9\textwidth]{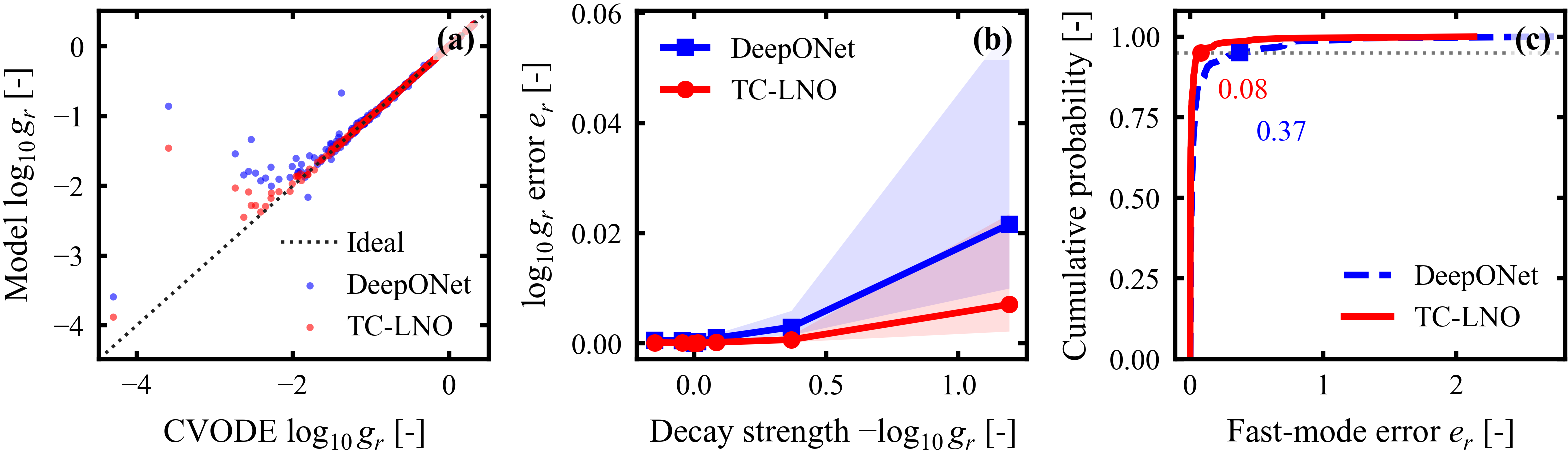}
    \caption{Finite-time modal response of TC-LNO and DeepONet for the DRM19 composition map over $\Delta t=2.00\times10^{-7}$~s. 
    Panel (a) compares the prediction-network and CVODE modal gains; panel (b) shows the median and interquartile range of the modal error as functions of the CVODE decay strength; 
    panel (c) shows the empirical cumulative distribution of $e_{M,r}$ for the 20\% most strongly attenuated retained modes. 
    The markers in panel (c) denote the $95^{th}$ percentile.}
    \label{fig:drm19_modal_response}
\end{figure*}

To characterize how the two operator networks represent locally different chemical responses, 
Fig.~\ref{fig:drm19_modal_response} analyzes the finite-time composition map over the deployed step $\Delta t=2\times10^{-7}$~s. 
The analysis uses the first 19 independent species in the componentwise standardized BCT space, while temperature and pressure are held fixed. 
Let $\boldsymbol{\Phi}_{M,\Delta t}$ denote the composition map of model $M\in\{\mathrm{CVODE},\mathrm{TC-LNO},\mathrm{DeepONet}\}$ and $\mathbf J_M$ its local Jacobian. 
The Jacobians are evaluated using centered perturbations of magnitude 0.002 at 64 validation states. 
Using the singular vectors of a finite-time linearized propagator to quantify perturbation amplification follows generalized and nonmodal stability analysis \cite{trefethenHydrodynamicStabilityEigenvalues1993,ducimetiereWeakNonlinearityStrong2022}. 
To place all models in the same modal basis, the CVODE Jacobian is decomposed as
\begin{equation}
    \mathbf J_{\mathrm{CVODE}}=\mathbf U\boldsymbol{\Sigma}\mathbf V^{\mathrm T}.
    \label{eq:cvode_jacobian_svd}
\end{equation}
In Eq.~\eqref{eq:cvode_jacobian_svd}, $\mathbf U$ and $\mathbf V$ contain the left and right singular vectors, respectively, 
and $\boldsymbol{\Sigma}$ is diagonal with singular values $\sigma_r$. The $r^{th}$ column $\mathbf v_r$ of $\mathbf V$, with $r=1,\ldots,19$, 
defines a common input direction for CVODE and both prediction networks. The finite-time gain of method $M$ along this direction is
\begin{equation}
    g_{M,r}=\left\lVert\mathbf J_M\mathbf v_r\right\rVert_2.
    \label{eq:modal_gain}
\end{equation}
For CVODE, this gain is the corresponding singular value $\sigma_r$. A gain below unity indicates attenuation of the perturbation over one step, whereas a gain above unity indicates amplification. The deviation of a predicted gain from the CVODE response is measured by
\begin{equation}
    e_{M,r}=\left|\log_{10}g_{M,r}-\log_{10}g_{\mathrm{CVODE},r}\right|.
    \label{eq:modal_gain_error}
\end{equation}
Thus, $e_{M,r}=0$ corresponds to exact agreement with the CVODE gain.  
A separate set of 19 modes is obtained at each validation state. The state index is omitted for compactness. 
Of the original 1,216 modes, those with $g_{\mathrm{CVODE},r}<10^{-5}$ are excluded because their finite-difference reference response is not robustly resolved, leaving 1,089 modes. 
The same conclusion is obtained for perturbation magnitudes from 0.001 to 0.004.

Figure~\ref{fig:drm19_modal_response}(a) shows that both networks reproduce the CVODE gains closely for the densely distributed modes near unity, 
whereas their separation becomes visible as the reference gain decreases. The TC-LNO points remain closer to the ideal line for most of these strongly attenuated modes, 
while several DeepONet points lie above it and therefore overestimate the retained response. Over all 1,089 retained modes, 
the median $e_{M,r}$ is $1.10\times10^{-4}$ for TC-LNO and $5.23\times10^{-4}$ for DeepONet, and TC-LNO gives the smaller error for 84.9\% of the modes. 
Figure~\ref{fig:drm19_modal_response}(b) groups the same errors by the CVODE decay strength $-\log_{10}g_{r}$. 
The medians of both networks are small in the weak-attenuation bins, but the median and interquartile range increase as the attenuation becomes stronger. 
Throughout this change, the TC-LNO curve and shaded interval remain below those of DeepONet. The strongest-attenuation subset in Fig.~\ref{fig:drm19_modal_response}(c) contains 218 modes, 
corresponding to 20\% of the retained modes. Its cumulative distribution rises more rapidly for TC-LNO, 
and the $95^{th}$-percentile errors are 0.081 for TC-LNO and 0.372 for DeepONet.
The smaller modal errors across the retained modes show that TC-LNO represents the finite-time gains of the DRM19 stiff chemistry map more accurately than DeepONet. 
This advantage is most evident for strongly attenuated modes, for which the TC-LNO gains remain closer to the CVODE reference.

\begin{figure}[!t]
    \centering
    \includegraphics[width=\columnwidth]{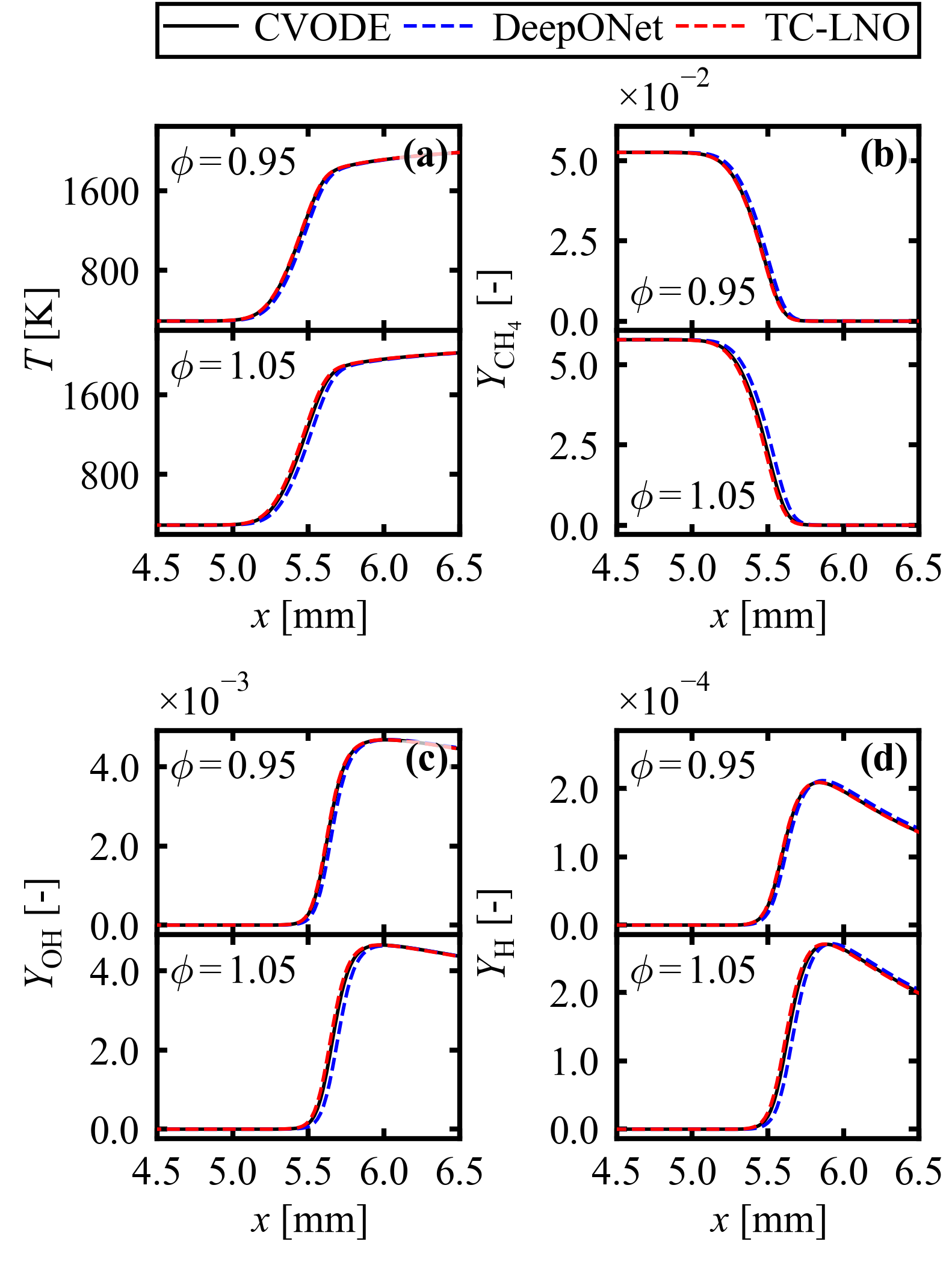}
    \caption{One-dimensional flame profiles predicted by CVODE, TC-LNO, and DeepONet baseline using DRM19.  
    Panels (a)--(d) show $T$, \ce{CH4}, OH, and H on the spatial coordinate.}
    \label{fig:drm19_lno_deeponet_profiles}
\end{figure}

The comparison of prediction performance is followed by the one-dimensional flame solution in Fig.~\ref{fig:drm19_lno_deeponet_profiles}. 
At both $\phi=0.95$ and $1.05$, DeepONet predicts a downstream shift of the reaction zone relative to CVODE, evident in the temperature and \ce{CH4} fronts as well as the 
OH and H profiles. In contrast, TC-LNO closely follows the reference through the reaction-zone onset, peak region, and downstream decay, 
accurately reproducing the position and magnitude of the radical profiles. 
These systematic spatial shifts in the DeepONet predictions account for the larger full-domain errors quantified in the next figure.

\begin{figure*}[!t]
    \centering
    \includegraphics[width=0.85\textwidth]{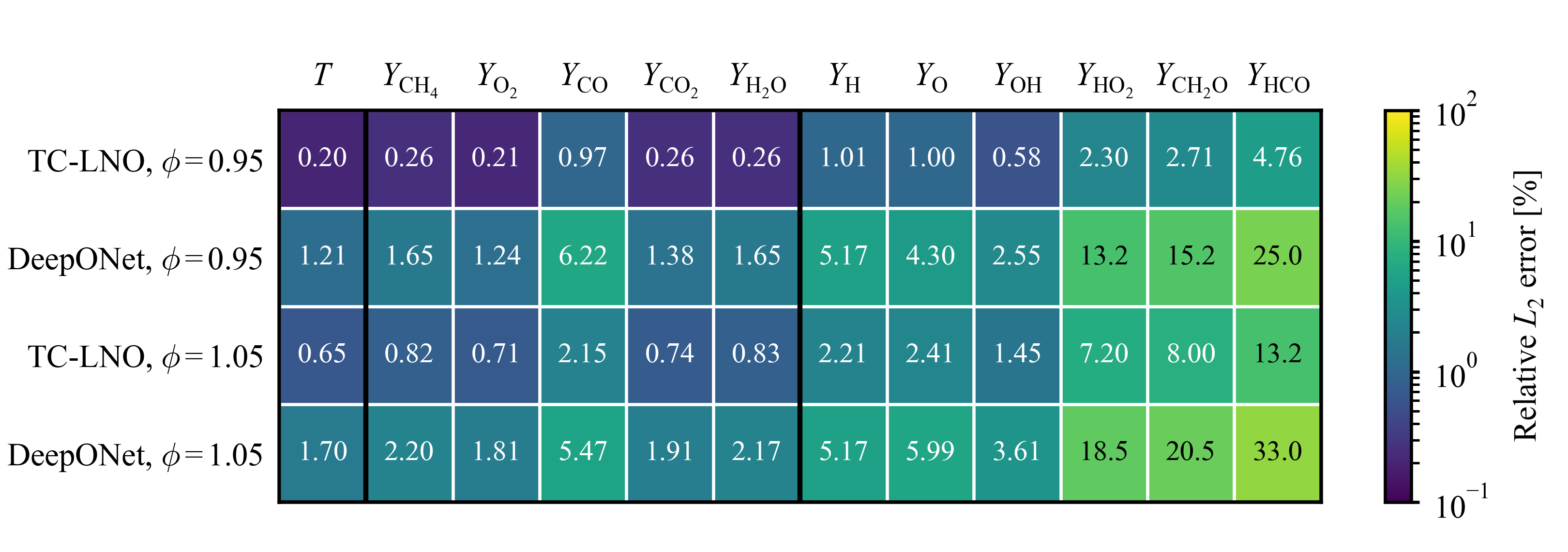}
    \caption{Relative spatial $L_2$ errors of temperature and representative species for TC-LNO and DeepONet at $\phi=0.95$ and 1.05.}
    \label{fig:drm19_lno_deeponet_errors}
\end{figure*}

Figure~\ref{fig:drm19_lno_deeponet_errors} quantifies the full-domain relative spatial $L_2$ errors, defined consistently with Eq.~\eqref{eq:sampling_macro_l2}. 
TC-LNO yields lower relative $L_2$ errors for all quantities at both equivalence ratios. At $\phi=0.95$, the relative $L_2$ errors for temperature and stable species are only 0.2\%--0.3\%, 
compared with 1.2\%--1.7\% for DeepONet, while the larger relative $L_2$ errors for low-concentration species are also substantially reduced. 
For example, the \ce{CH2O} and \ce{HCO} errors decrease from 15.2\% and 25.0\% to 2.7\% and 4.8\%, respectively. 
The species macro-average consequently decreases from 13.0\% for DeepONet to 2.3\% for TC-LNO. 
At $\phi=1.05$, the relative $L_2$ errors increase for both networks, consistent with the larger profile displacement in Fig.~\ref{fig:drm19_lno_deeponet_profiles}, 
but TC-LNO remains more accurate for every quantity, reducing the \ce{CH2O} and \ce{HCO} errors from 20.5\% and 33.0\% to 8.0\% and 13.2\%, respectively, 
and the macro-average from 17.1\% to 6.7\%. Figs.~\ref{fig:drm19_modal_response}--\ref{fig:drm19_lno_deeponet_errors} collectively demonstrate that, 
under the same data and training configuration, TC-LNO represents the stiff chemistry map more accurately than the implemented DeepONet baseline.

\subsubsection{TC-MLNO compared with the ungrouped TC-LNO}
\label{subsubsec:mlno_lno_results}

\begin{figure}[!t]
    \centering
    \includegraphics[width=\columnwidth]{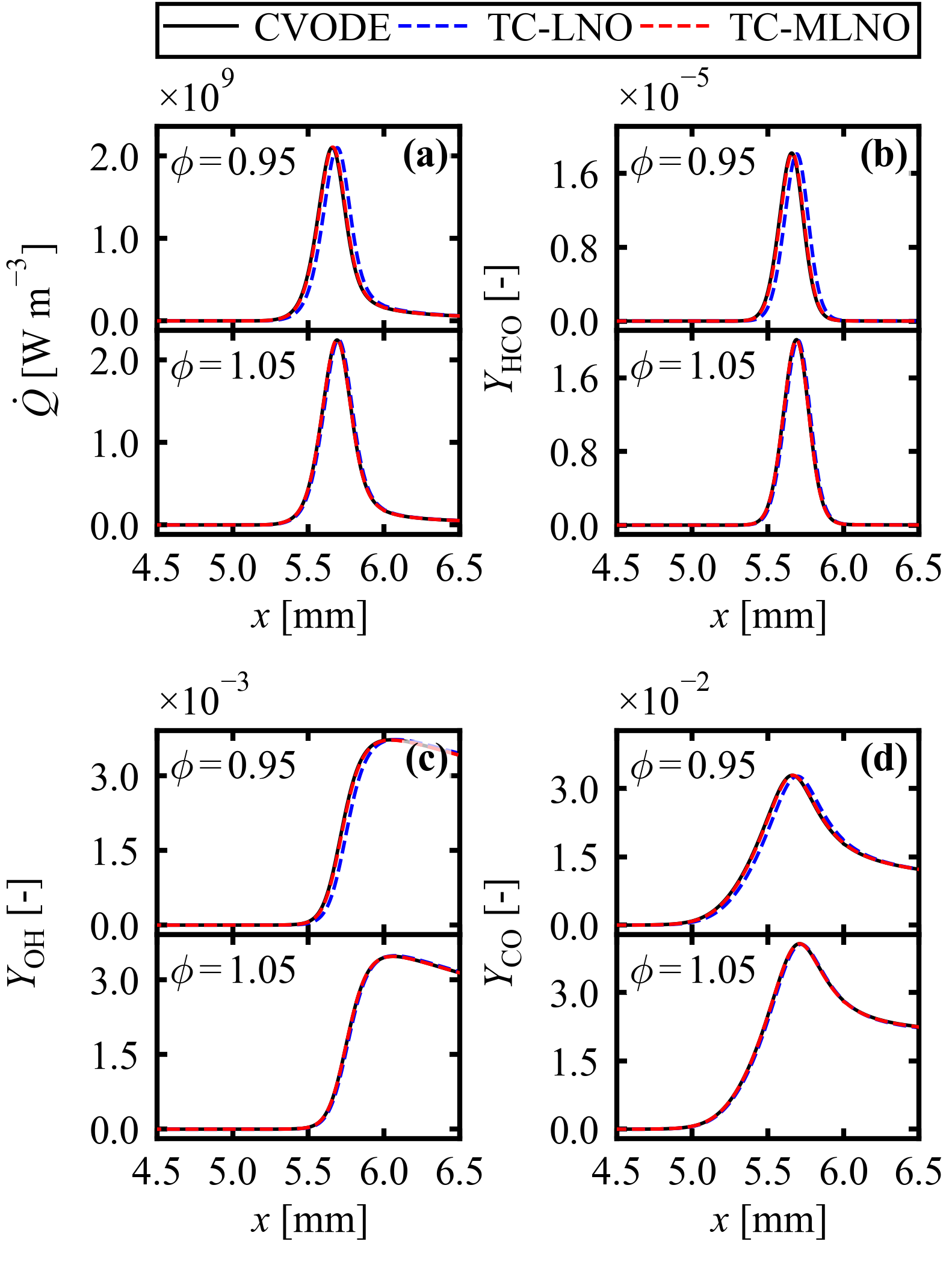}
    \caption{One-dimensional flame profiles predicted by CVODE, TC-LNO, and TC-MLNO using GRI-Mech~3.0. 
    Panels (a)--(d) show heat release rate $\dot Q$, \ce{HCO}, OH, and CO.}
    \label{fig:gri3_lno_mlno_profiles}
\end{figure}

The larger GRI-Mech~3.0 mechanism introduces both a higher-dimensional composition space and a substantially wider range of reaction time scales. 
Following the procedure in Section~\ref{subsubsec:timescale_grouping}, the 52 predicted species without inert species are divided into fast, medium, and slow groups containing 16, 16, and 20 species, respectively. 
The detailed timescale ordering and group boundaries are provided in Fig~SC.~2 of the Supplementary C.

Figure~\ref{fig:gri3_lno_mlno_profiles} compares the ungrouped TC-LNO and reaction-timescale-grouped TC-MLNO trained on the same one-million-state conditional diffusion dataset. 
Both reproduce the overall structure of species and heat release rate $\dot Q$ profiles at both equivalence ratios. 
At $\phi=0.95$, however, the TC-LNO predicts a downstream shift of the reaction zone relative to CVODE, reflected in
the $\dot Q$, \ce{HCO}, OH, and CO profiles. TC-MLNO more accurately captures their positions and shapes and remains closer to CVODE throughout the reaction and post-flame regions. 
At $\phi=1.05$, the three solutions agree more closely, with the remaining differences concentrated near the peaks and steep gradients, where TC-MLNO remains in closer agreement.

\begin{figure}[!t]
    \centering
    \includegraphics[width=\columnwidth]{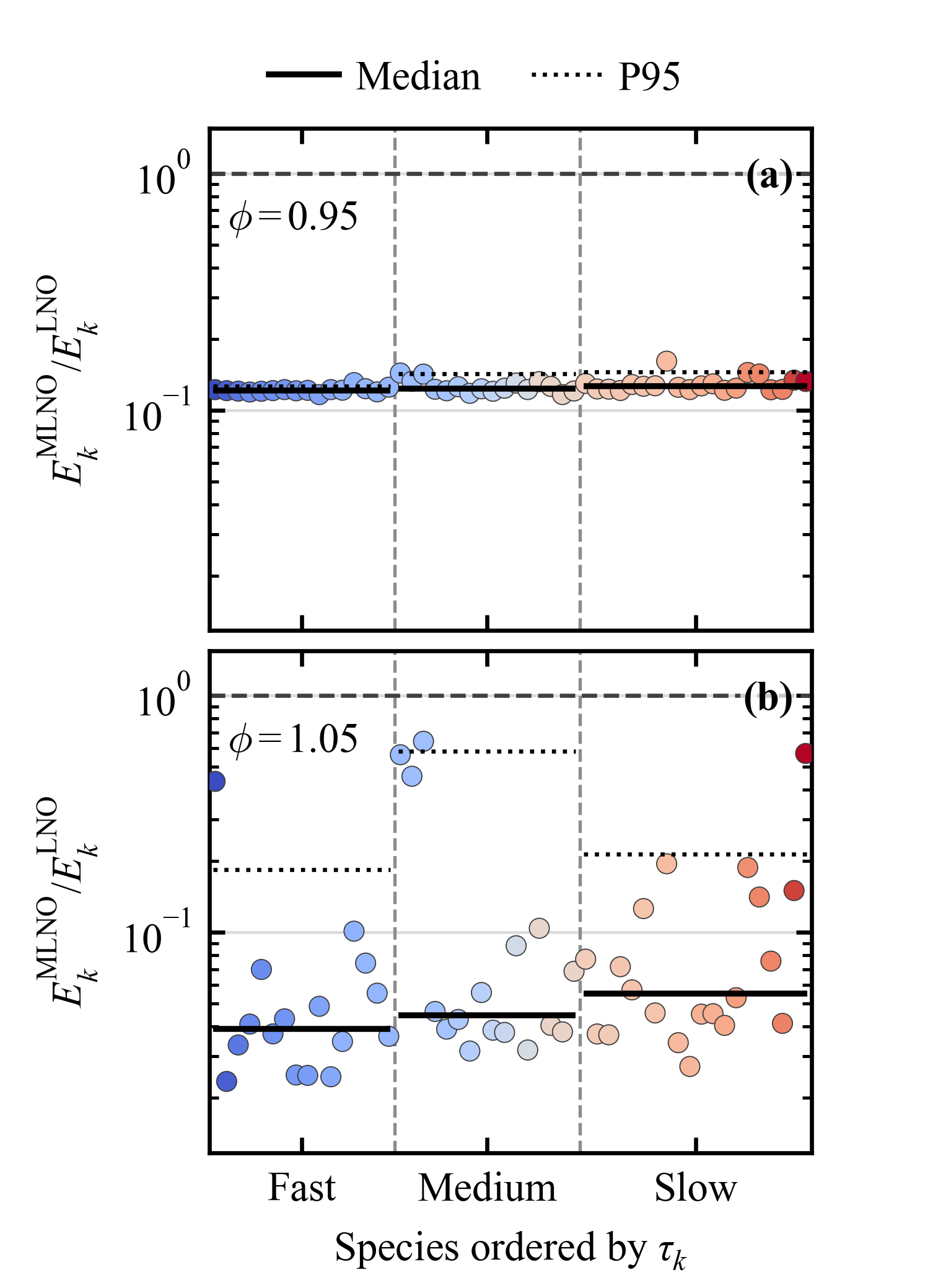}
    \caption{Species-wise ratio of TC-MLNO and TC-LNO relative spatial $L_2$ errors at (a) $\phi=0.95$ and (b) $\phi=1.05$. 
    The solid and dotted lines give the median and $95^{th}$ percentile within each time-scale group.}
    \label{fig:gri3_lno_mlno_error_ratio}
\end{figure}

Figure~\ref{fig:gri3_lno_mlno_error_ratio} compares all species using the error ratio $E_k^{\mathrm{MLNO}}/E_k^{\mathrm{LNO}}$, where values below unity favor TC-MLNO.
At $\phi=0.95$, the median relative $L_2$ errors decrease from 32.3\%, 18.5\%, and 14.2\% for the fast, medium, and slow groups to 3.9\%, 2.3\%, and 1.7\%, respectively.
The corresponding median error ratios are 0.121, 0.124, and 0.126, and the $95^{th}$-percentile ratios remain below 0.15 for all three groups.
At $\phi=1.05$, the median errors decrease from 8.6\%, 5.2\%, and 4.1\% to 0.3\%, 0.3\%, and 0.2\%, respectively, with median error ratios of 0.039, 0.045, and 0.055.
Although the degree of improvement varies among species, particularly within the medium group at $\phi=1.05$, TC-MLNO yields a lower relative $L_2$ error than TC-LNO for all 52 predicted species at both equivalence ratios.
Thus, the improvement extends across the fast, medium, and slow groups rather than being driven by a small subset of species.
Figures~\ref{fig:gri3_lno_mlno_profiles}--\ref{fig:gri3_lno_mlno_error_ratio} establish that the Laplace neural operator remains effective for detailed kinetics and 
that the reaction-timescale-grouped TC-MLNO substantially improves the a posteriori predictions of the ungrouped TC-LNO across the three species groups.

\subsection{Two-dimensional temporally evolving jet flame}
\label{subsec:2djet_flame}

To assess the chemistry prediction networks under shear deformation and turbulence--chemistry coupling, a two-dimensional temporally evolving jet is considered.
The $16\times16$~mm$^2$ domain contains a 1~mm-wide unburned premixed stream centered at $y=8$~mm, surrounded by the downstream burned state of the corresponding flame.
The premixed gas is at $\phi=1.0$, $T_0=300$~K, and $P_0=101325$~Pa, with fuel mole fractions $X_{\mathrm{NH_3}}:X_{\mathrm{H_2}}:X_{\mathrm{N_2}}=0.40:0.45:0.15$.
Following the double-tanh jet profile and divergence-free multimode perturbation of Gourianov et al.~\cite{gourianovQuantuminspiredApproachExploit2022}, 
the initial velocity is written as
\begin{equation}
    \mathbf{u}(x,y,0)=U(y)\mathbf{e}_x+\mathbf{u}'(x,y),
    \label{eq:jet_initial_velocity}
\end{equation}
\begin{equation}
\begin{split}
    U(y) &= U_{\mathrm{o}}+\frac{U_{\mathrm{j}}-U_{\mathrm{o}}}{2}
    \left[\tanh\!\left(\frac{y-y_1}{h}\right)\right. \\
    &\quad \left.-\tanh\!\left(\frac{y-y_2}{h}\right)\right],
\end{split}
\label{eq:jet_double_tanh}
\end{equation}
where $\mathbf{e}_x$ is the streamwise unit vector, $U_{\mathrm{j}}=33.58$ and $U_{\mathrm{o}}=-84$~m\,s$^{-1}$,
$y_1=7.5$~mm, $y_2=8.5$~mm, and $h=0.08$~mm.
The perturbation combines streamwise modes $n=3$, 4, and 12, with wave numbers $2\pi n/L_x$ for $L_x=16$~mm,
and is normalized to a maximum magnitude of 2.94~m\,s$^{-1}$.

The $800\times550$ mesh is uniform streamwise and refined across the jet and shear layers.
The solver uses mixture-averaged transport and C3MechLite \ce{H2}/\ce{NH3} mechanism (24 species, 165 reactions) \cite{murakamiC3MechLiteIntegratedComponent2025}.
To compare the two operator architectures under the same multiscale treatment of chemistry, the DeepONet baseline
also adopts the fast, medium, and slow species grouping and is denoted multiscale DeepONet (MDeepONet). Both models
predict 9, 7, and 7 species in the respective groups and share the conditional-diffusion training data, BCT
preprocessing, and physical-loss weights.

\begin{figure}[!t]
    \centering
    \includegraphics[width=\columnwidth]{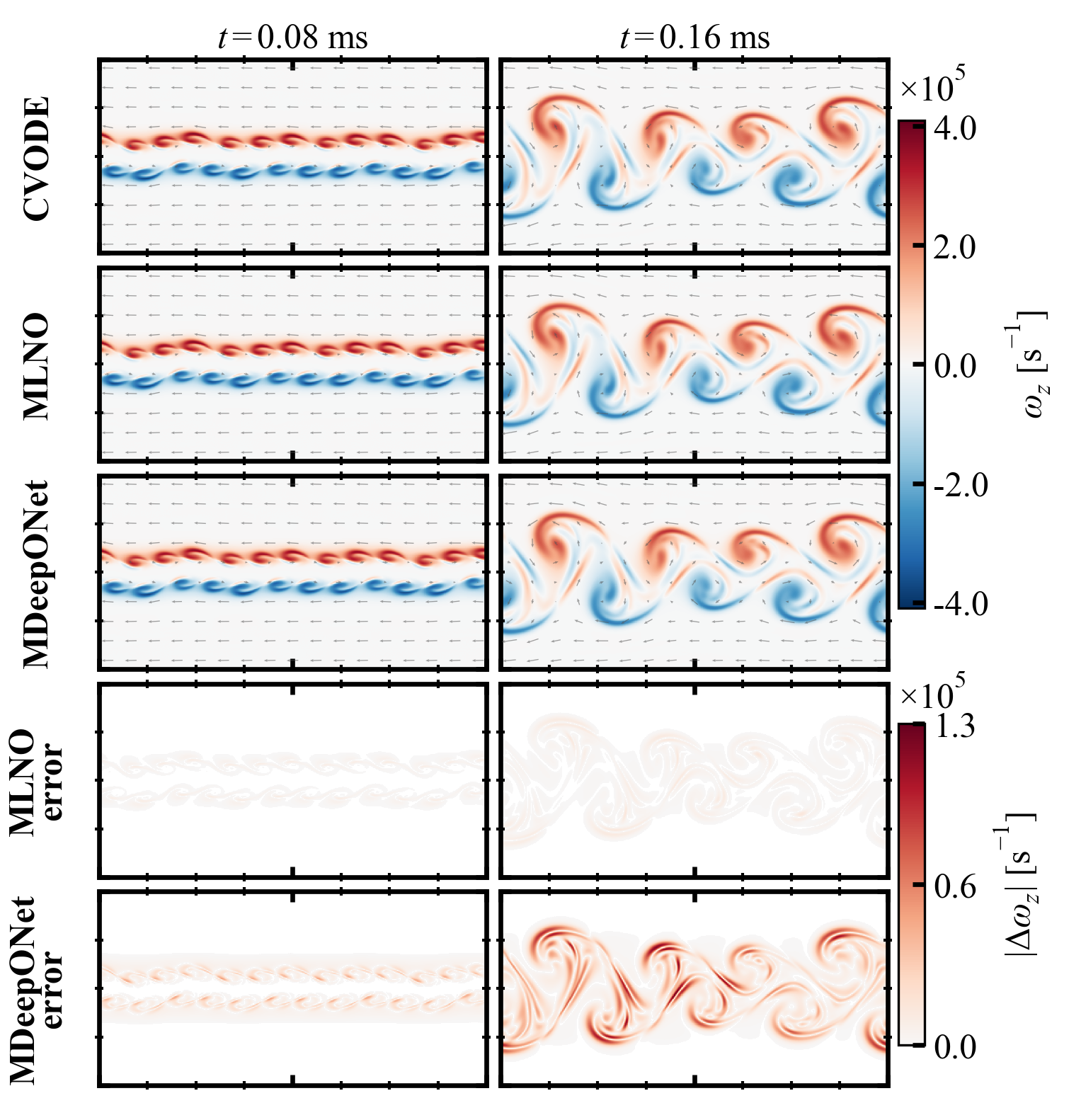}
    \caption{Vorticity ($\omega_z$) evolution of the two-dimensional double-tanh jet flame at $t=0.08$ and 0.16~ms.}
    \label{fig:2djet_qdot_evolution}
\end{figure}

Figure~\ref{fig:2djet_qdot_evolution} compares the vorticity fields at $t=0.08$ and 0.16~ms.
The CVODE solution changes from two nearly planar, oppositely signed vorticity layers to corrugated shear layers with pronounced roll-up and spiral-shaped vortices. 
TC-MLNO follows this evolution closely, retaining the positions and shapes of the principal vortical structures.
MDeepONet reproduces the broad sequence of deformation but develops larger discrepancies in vorticity intensity and vortex position as the shear layers roll up.
The absolute-error maps show that the TC-MLNO discrepancies remain smaller,
whereas the MDeepONet errors extend along substantial portions of the rolled-up shear layers.
Quantitative assessment uses the single-variable relative $L_2$ error underlying Eq.~\eqref{eq:sampling_macro_l2} for vorticity. 
The relative $L_2$ error is evaluated using volume-weighted averages over the full domain.
Unlike the absolute-error maps, which show pointwise deviations from CVODE, 
the relative $L_2$ error quantifies the overall full-domain discrepancy for each variable separately.
At the $t=0.08$ and 0.16~ms, the TC-MLNO relative $L_2$ errors in vorticity are 1.3\% and 2.5\%,
compared with 6.4\% and 24.3\% for MDeepONet.
Thus, TC-MLNO retains closer agreement with the coupled reference flow as the vortical structures become more complex, with the vorticity error at $t=0.16$~ms reduced by approximately a factor of 9.5.
The closer agreement during repeated transport--chemistry coupling is qualitatively consistent with the more accurate finite-time attenuation of perturbations in
Fig.~\ref{fig:drm19_modal_response}.

\begin{figure}[!t]
    \centering
    \includegraphics[width=\columnwidth]{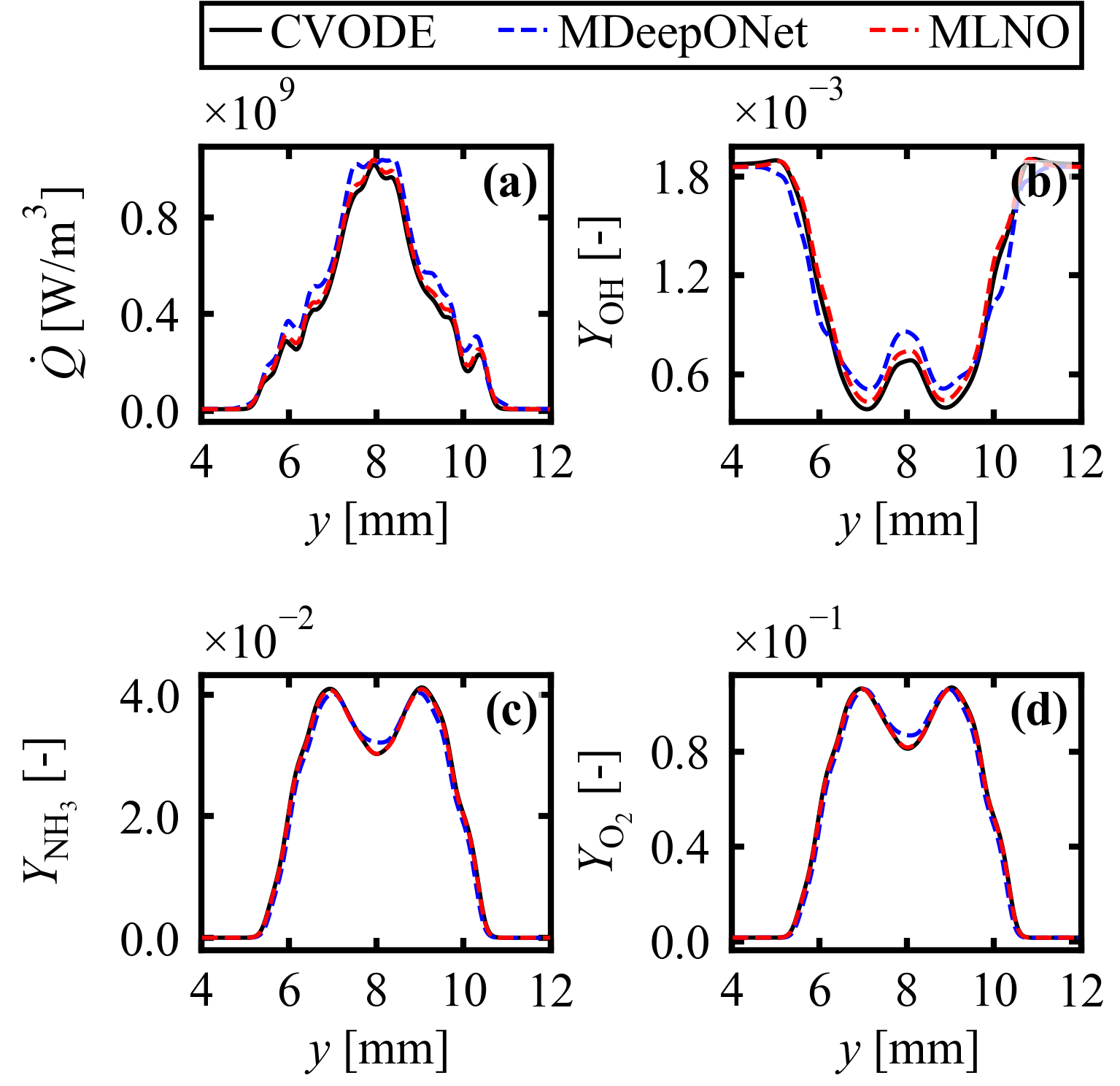}
    \caption{Instantaneous streamwise-averaged transverse profiles at $t=0.16$~ms:
    (a) $\dot Q$, and mass fractions of (b) OH, (c) \ce{NH3}, and (d) \ce{O2}.}
    \label{fig:2djet_statistical_profiles}
\end{figure}

Figure~\ref{fig:2djet_statistical_profiles} further compares the transverse profiles at $t=0.16$~ms.
These are instantaneous streamwise averages evaluated using cell-volume weights within each mesh row.
The mean $\dot Q$ is largest near the center of the deformed jet.
TC-MLNO resolves the central peak $\dot Q$ and its lateral shoulders.
MDeepONet overpredicts $\dot Q$ over much of the reacting region.
The latter discrepancy remains visible after streamwise averaging, indicating a systematic intensity error in addition to the local front displacements.
The species profiles provide a complementary assessment.
TC-MLNO captures the two off-center maxima and the central depression of \ce{NH3} and \ce{O2},
whereas MDeepONet predicts shallower central depressions and larger deviations along the outer gradients.
The distinction is more pronounced for OH. MDeepONet overpredicts the central peak and the adjacent minimum values,
while TC-MLNO follows the CVODE profile more closely, with a smaller residual overprediction in the inner region.
The combined instantaneous and averaged comparisons show that TC-MLNO more accurately preserves the evolving vortical structures,
heat-release intensity, and the distributions of both radicals and reactants throughout the simulated jet evolution.

\subsection{Three-dimensional Darmstadt burner flame}
\label{subsec:darmstadt_flame}

\begin{figure}[!t]
    \centering
    \includegraphics[width=0.8\columnwidth]{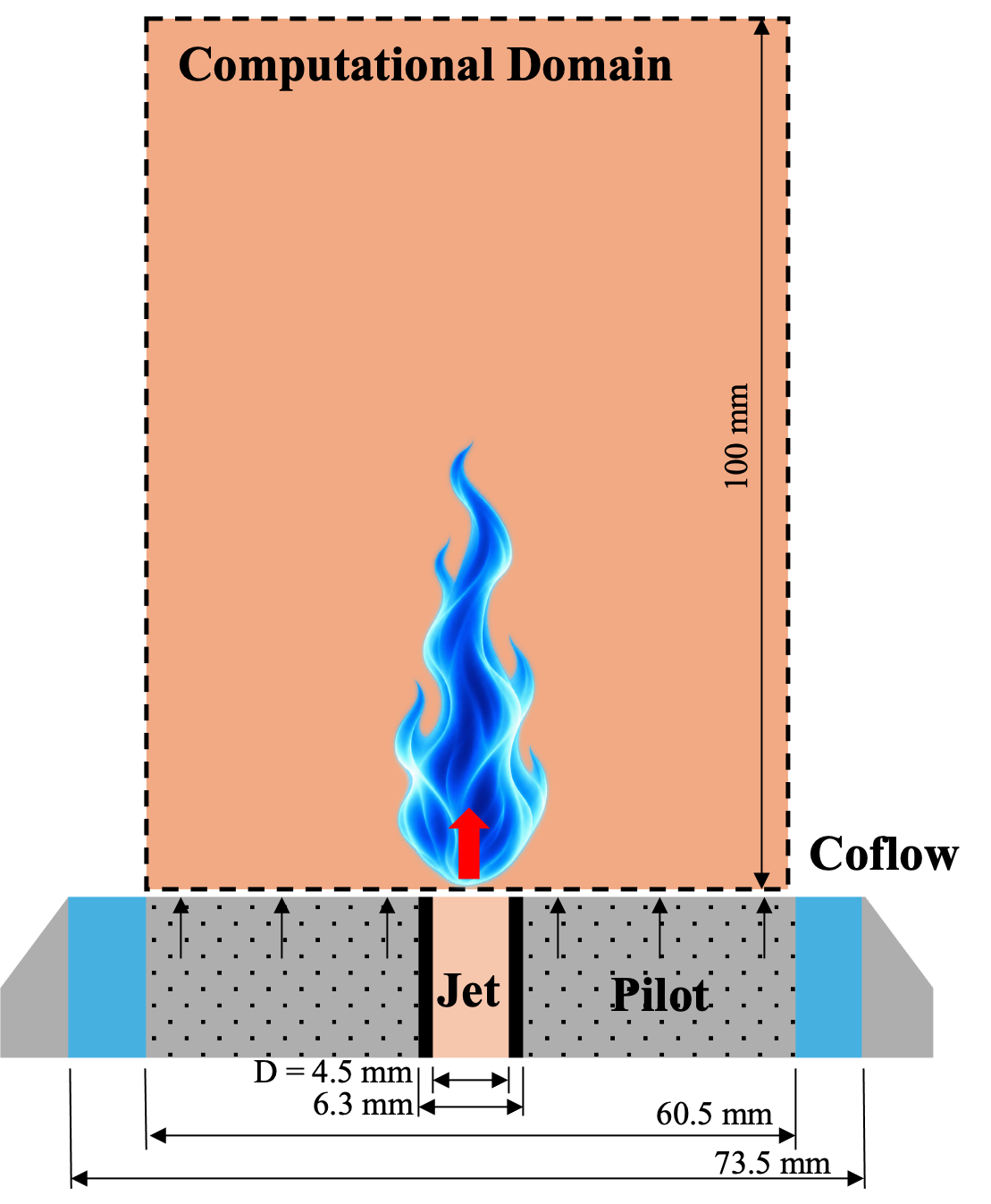}
    \caption{Schematic of the three-dimensional Darmstadt burner configuration and the computational domain.}
    \label{fig:darmstadt_domain}
\end{figure}

The three-dimensional Darmstadt piloted premixed jet flame is considered to further assess the chemistry prediction
networks under turbulent transport and three-dimensional flame deformation \cite{schultheisQuantitativeMeasurementsThermochemical2024}.
The configuration consists of a central lean premixed jet surrounded by a burned-gas pilot, with a nozzle diameter
of $D=4.5$~mm, a bulk jet velocity of 70~m/s, and an equivalence ratio of $\phi=0.57$.
The ammonia-hydrogen fuel has mole fractions $X_{\mathrm{NH_3}}:X_{\mathrm{H_2}}:X_{\mathrm{N_2}}
=0.40:0.45:0.15$ and is premixed with air before entering at 300~K and atmospheric pressure.
This configuration introduces three-dimensional turbulent entrainment and interactions between the cold jet and the surrounding pilot products.

Figure~\ref{fig:darmstadt_domain} shows the computational configuration.
The cylindrical domain extends 100~mm downstream and has an outer diameter of 73.5~mm.
The numerical inlet retains the central jet and the separating lip up to a diameter of 6.3~mm.
The surrounding annulus is represented by a uniform burned-gas pilot extending to the radius of 30.25~mm.
The prescribed pilot velocity is 2.5~m/s, with fixed burned-gas species mass fractions.
These settings define a simplified numerical realization of the burner for the CVODE--network comparison.
Large-eddy simulation is performed with a Smagorinsky subgrid model on a common mesh of $7.1\times10^6$ cells.
Mixture-averaged transport and the C3MechLite mechanism are used.
The flow equations use first-order implicit time integration, upwind momentum convection, and limited linear scalar convection.
TC-MLNO and MDeepONet use the same training dataset and normalization, the same 9/7/7 species partition, and identical physical-loss weights.
The simulation is first advanced on a coarse mesh to $t=5$~ms, after which the solution is mapped onto the refined mesh. Starting from this state, 
the simulations using TC-MLNO and MDeepONet are each advanced for an additional 2000 time steps.

\begin{figure}[!t]
    \centering
    \includegraphics[width=\columnwidth]{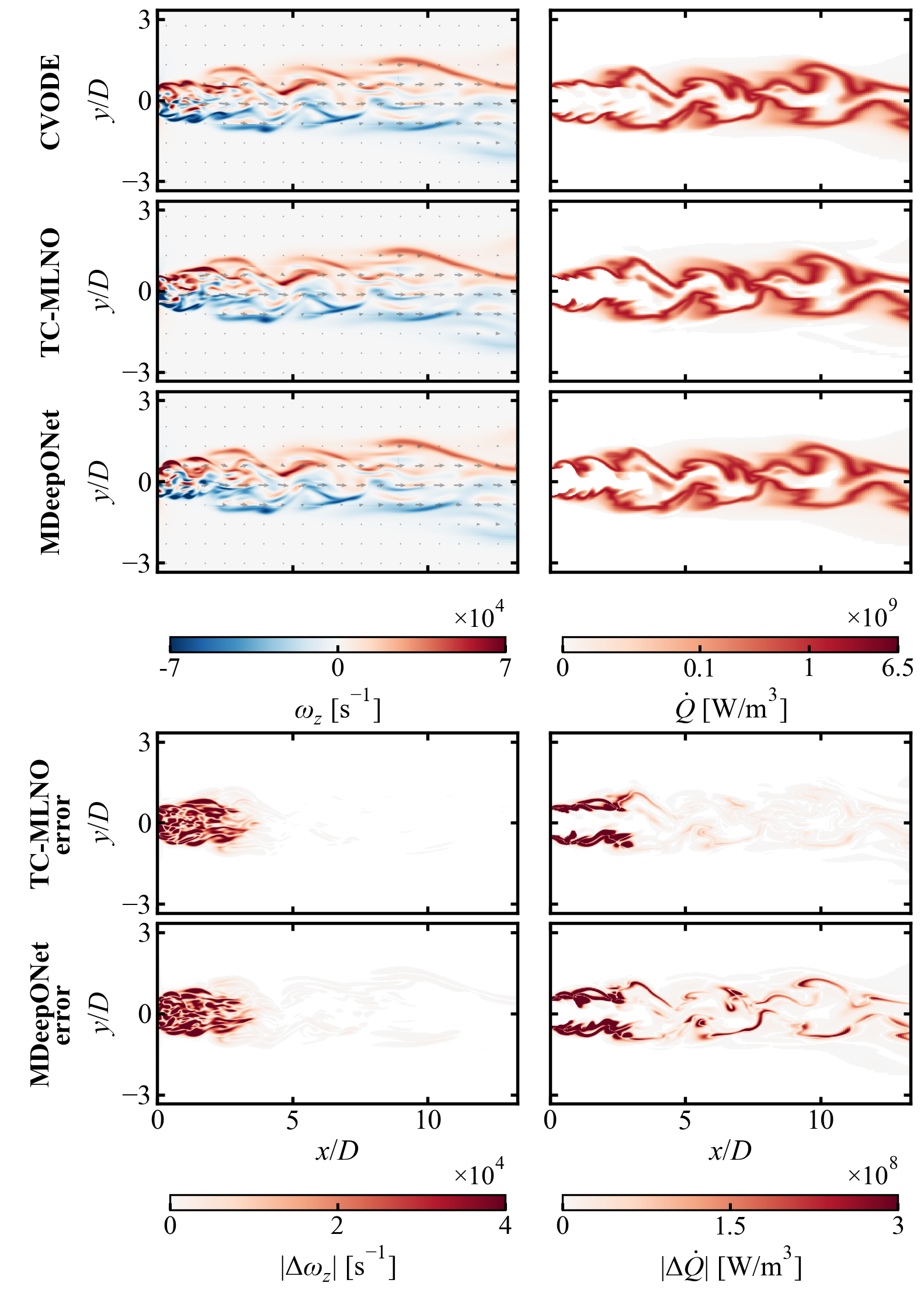}
    \caption{Instantaneous center-plane distributions of vorticity ($\omega_z$) and $\dot Q$.
    The displayed region is normalized by $D=4.5$~mm.}
    \label{fig:darmstadt_instantaneous}
\end{figure}

Figure~\ref{fig:darmstadt_instantaneous} compares the center-plane distributions of vorticity and $\dot Q$.
All three solutions reproduce the principal shear-layer structures and the surrounding folded heat-release layers.
The vorticity fields show the downstream deflection and broadening of the jet, while heat release indicates the deformed reacting layers.
Both operator networks follow the principal structures of the CVODE solution.
The absolute vorticity errors of both models are largest near the nozzle, while TC-MLNO gives smaller heat-release discrepancies along much of the downstream reacting layers.
The corresponding differences in reactive-species predictions are examined next through radial profiles and overall relative $L_2$ errors.

\begin{figure}[!t]
    \centering
    \includegraphics[width=\columnwidth]{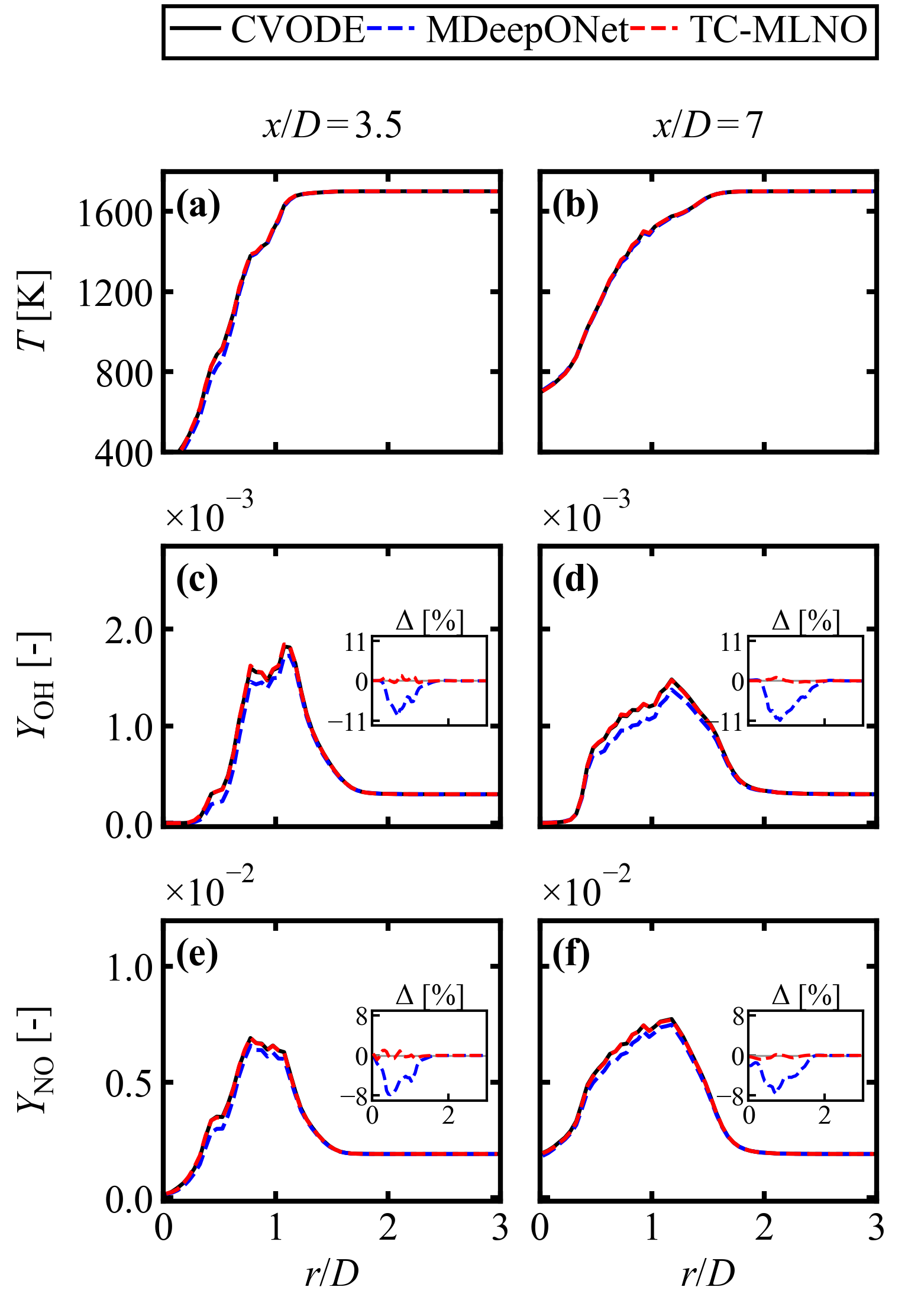}
    \caption{Instantaneous volume-weighted radial profiles of temperature, OH, and NO at $x/D=3.5$, 7, and 10.5.
    Insets show the signed deviation $\Delta=100(f_{\mathrm{model}}-f_{\mathrm{CVODE}})/(f_{\mathrm{CVODE,max}}-f_{\mathrm{CVODE,min}})$.}
    \label{fig:darmstadt_radial}
\end{figure}

Figure~\ref{fig:darmstadt_radial} presents instantaneous, volume-weighted radial profiles of temperature, OH, and NO at $x/D=3.5$ and 7.
Both models reproduce the overall temperature rise from the jet interior to the surrounding hot gas. 
However, MDeepONet develops a visible offset through the inner transition at the upstream station, whereas TC-MLNO remains close to CVODE at both locations.
The accumulated difference is substantially clearer in the reactive-species profiles.
Both OH and NO exhibit off-axis maxima within the reacting region and then decay toward the pilot background.
MDeepONet systematically underpredicts their peaks and shoulders, while TC-MLNO closely tracks both the magnitude and radial position of the CVODE profiles.
Consistent with the profile comparison, the inset deviations of MDeepONet are predominantly negative and extend across
the peak and shoulder regions, indicating a systematic underprediction of the reactive-species levels.
In contrast, the TC-MLNO deviations remain narrowly distributed around zero and rapidly diminish outside the reacting region.
At the two stations, the maximum absolute deviations of the OH profiles are 9.6\% and 10.9\% for MDeepONet, 
compared with only 1.5\% and 0.9\% for TC-MLNO.
For NO, the corresponding deviations decrease from 7.8\% and 7.6\% to 1.0\% and 0.8\%.
The advantage of TC-MLNO is thus sustained from the fast-group radical OH to the slow-group species NO and across both axial locations.
Over the entire sampled slab, the volume-weighted relative $L_2$ errors of OH, and NO decrease 
from 11.4\% and 5.5\% for MDeepONet to 1.4\% and 0.6\% for TC-MLNO, corresponding to reductions of 88.0\% and 88.9\%, respectively.
Figures~\ref{fig:darmstadt_instantaneous}--\ref{fig:darmstadt_radial} show that 
TC-MLNO continues to preserve the principal three-dimensional flame structure and reactive-species distributions, 
whereas MDeepONet accumulates substantially larger errors in the radical and NO mass fractions.

The computational cost is assessed using calculations performed with 32 CPU processes on a dual-socket AMD EPYC 7302 system and 6 NVIDIA GeForce RTX 4090 GPUs. 
Table~\ref{tab:darmstadt_computational_cost} reports the chemistry time ($t_{\mathrm{chem}}$) and the total time ($t_{\mathrm{tot}}$) per step. 
Relative to CVODE, TC-MLNO reduces the mean chemistry time by 90.1\% and the mean total time by 62.6\%, while the corresponding reductions for MDeepONet are 92.9\% and 60.7\%.
Assuming unchanged mean costs, linear extrapolation to a 5~ms simulation gives estimated savings of approximately 323~h for TC-MLNO and 313~h for MDeepONet relative to CVODE.
The TC-MLNO training cost of approximately 12~h is about 3.7\% of its projected saving.
Therefore, TC-MLNO provides the higher long-time prediction accuracy while retaining a total computational cost comparable to 
MDeepONet and substantially reducing the expense of direct chemical integration with CVODE.

\begin{table}[!t]
    \centering
    \caption{Computational costs for CVODE (32 cores) and operator networks (6 GPU cards).}
    \label{tab:darmstadt_computational_cost}
    \fontsize{8pt}{9pt}\selectfont
    \setlength{\tabcolsep}{3pt}
    \begin{tabular}{l c c c}
        \hline
        Method 
        & \shortstack{Chemistry time\\$t_{\mathrm{chem}}$ (s)}
        & \shortstack{Total time\\$t_{\mathrm{tot}}$ (s)}
        & \shortstack{$t_{\mathrm{chem}}/t_{\mathrm{tot}}$\\(\%)} \\
        \hline
        CVODE     & 47.03 & 74.33 & 63.3 \\
        TC-MLNO      & 4.68  & 27.82 & 16.8 \\
        MDeepONet & 3.36  & 29.22 & 11.5 \\
        \hline
    \end{tabular}
\end{table}

\section{Conclusions}
\label{sec:Conclusions}

This work develops an integrated framework for accelerating stiff chemical kinetics by coupling a conditional diffusion sampling strategy with a thermochemically constrained multiscale Laplace neural operator (TC-MLNO).
The diffusion model learns the joint distribution of temperature and species conditioned on equivalence ratio, allowing low-dimensional flame data to be augmented without relying on
independently prescribed perturbations. Compared with random sampling, the resulting states more closely follow the correlated variations of fuel consumption, temperature rise,
and radical formation. This improved representation of the reacting-state distribution consistently reduces the a posteriori errors in coupled flame calculations,
with the greatest gains observed for high-dimensional chemical states, minor species, and radicals.

TC-MLNO complements the improved training data by combining pole--residue Laplace layers and reaction-timescale-specific branches with thermochemical constraints.
Across one-dimensional flames and multidimensional reacting-flow simulations,
TC-MLNO more accurately preserves reaction-zone locations, heat-release structures, and the distributions of chemically sensitive species than the comparison operator.
Its advantage remains evident during repeated transport--chemistry coupling, where reduced local errors lead to markedly lower long-time discrepancies in radical
and nitrogen-containing species. In the three-dimensional turbulent flame, the OH and NO errors are reduced by 88.0\% and 88.9\%, respectively. Relative to direct stiff integration,
TC-MLNO reduces the mean chemistry time by approximately 90.0\% and the total solution time by approximately 63.0\%,
demonstrating a favorable balance between predictive accuracy and computational efficiency.

Overall, conditional diffusion sampling and TC-MLNO address the data-distribution and dynamics-representation challenges of learning stiff chemical ODEs,
providing an accurate and efficient alternative to repeated chemical integration in multidimensional reacting-flow simulations.
Future work extends the framework to broader fuels and operating conditions with flexible chemical time increments.

\section*{CrediT authorship contribution statement}

\textbf{Dong Yang}: Writing - original draft, Software, Investigation, Visualization.
\textbf{Xiongbo Duan}: Writing - review \& editing, Methodology.
\textbf{Min Zhang}: Writing - review \& editing, Software, Investigation, Methodology, Conceptualization.
\textbf{Yuqiang Li}: Writing - review \& editing, Supervision.

\section*{Declaration of competing interest}

The authors declare that they have no known competing financial interests or personal relationships that could have appeared to influence the work reported in this paper.

\section*{Acknowledgments}

This work is supported by the National Natural Science Foundation of China (Grant No. 52506177). 
We are grateful for the resources provided by the High-Performance Computing Center of Central South University.

\section*{Supplementary material}

Supplementary material associated with this article can be found, in the online version.

\FloatBarrier

\bibliographystyle{cnf-num}
\bibliography{LaplaceNet}

@article{anExploreArtificialNeural2022,
  title = {Explore Artificial Neural Networks for Solving Complex Hydrocarbon Chemistry in Turbulent Reactive Flows},
  author = {An, Jian and Qin, Fei and Zhang, Jian and Ren, Zhuyin},
  year = 2022,
  month = jul,
  journal = {Fundam. Res.},
  volume = {2},
  number = {4},
  pages = {595--603},
  issn = {2667-3258},
  doi = {10.1016/j.fmre.2021.08.007},
  langid = {english}
}

@article{azizzadenesheliNeuralOperatorsAccelerating2024,
  title = {Neural Operators for Accelerating Scientific Simulations and Design},
  author = {Azizzadenesheli, Kamyar and Kovachki, Nikola and Li, Zongyi and {Liu-Schiaffini}, Miguel and Kossaifi, Jean and Anandkumar, Anima},
  year = 2024,
  month = may,
  journal = {Nat. Rev. Phys.},
  volume = {6},
  number = {5},
  pages = {320--328},
  issn = {2522-5820},
  doi = {10.1038/s42254-024-00712-5},
  langid = {english}
}

@article{boxAnalysisTransformations1964,
  title = {An Analysis of Transformations},
  author = {Box, G. E. P. and Cox, D. R.},
  year = 1964,
  month = jul,
  journal = {J. R. Stat. Soc. Ser. B Methodol.},
  volume = {26},
  number = {2},
  pages = {211--243},
  issn = {0035-9246},
  doi = {10.1111/j.2517-6161.1964.tb00553.x},
  urldate = {2026-08-28},
  langid = {english}
}

@article{caoLaplaceNeuralOperator2024,
  title = {Laplace Neural Operator for Solving Differential Equations},
  author = {Cao, Qianying and Goswami, Somdatta and Karniadakis, George Em},
  year = 2024,
  month = jun,
  journal = {Nat. Mach. Intell.},
  volume = {6},
  number = {6},
  pages = {631--640},
  issn = {2522-5839},
  doi = {10.1038/s42256-024-00844-4},
  urldate = {2026-05-16},
  langid = {english}
}

@article{chiOntheflyArtificialNeural2021,
  title = {On-the-Fly Artificial Neural Network for Chemical Kinetics in Direct Numerical Simulations of Premixed Combustion},
  author = {Chi, Cheng and Janiga, G{\'a}bor and Th{\'e}venin, Dominique},
  year = 2021,
  month = apr,
  journal = {Combust. Flame},
  volume = {226},
  pages = {467--477},
  issn = {0010-2180},
  doi = {10.1016/j.combustflame.2020.12.038},
  langid = {english}
}

@article{christoArtificialNeuralNetwork1996,
  title = {Artificial Neural Network Implementation of Chemistry with Pdf Simulation of {{\ce{H2}}}/{{\ce{CO2}}} Flames},
  author = {Christo, F. C. and Masri, A. R. and Nebot, E. M.},
  year = 1996,
  month = sep,
  journal = {Combust. Flame},
  volume = {106},
  number = {4},
  pages = {406--427},
  issn = {0010-2180},
  doi = {10.1016/0010-2180(95)00250-2},
  langid = {english}
}

@article{dingMachineLearningTabulation2021,
  title = {Machine Learning Tabulation of Thermochemistry in Turbulent Combustion: {{An}} Approach Based on Hybrid Flamelet/Random Data and Multiple Multilayer Perceptrons},
  shorttitle = {Machine Learning Tabulation of Thermochemistry in Turbulent Combustion},
  author = {Ding, Tianjie and Readshaw, Thomas and Rigopoulos, Stelios and Jones, W. P.},
  year = 2021,
  month = sep,
  journal = {Combust. Flame},
  volume = {231},
  pages = {111493},
  issn = {00102180},
  doi = {10.1016/j.combustflame.2021.111493},
  urldate = {2024-12-22},
  langid = {english}
}

@article{ducimetiereWeakNonlinearityStrong2022,
  title = {Weak Nonlinearity for Strong Non-Normality},
  author = {Ducimeti{\`e}re, Yves-Marie and Boujo, Edouard and Gallaire, Fran{\c c}ois},
  year = 2022,
  journal = {J. Fluid Mech.},
  edition = {2022/08/31},
  volume = {947},
  pages = {A43},
  publisher = {Cambridge University Press},
  issn = {0022-1120},
  doi = {10.1017/jfm.2022.664},
  langid = {english}
}

@article{gcerri.KineticCombustionNeural2003,
  title = {Kinetic Combustion Neural Modelling Integrated into Computational Fluid Dynamics},
  author = {Cerri, G. and Michelassi, V. and Monacchia, S. and Pica, S.},
  year = 2003,
  month = jan,
  journal = {Proc. Inst. Mech. Eng. Part A-J. Power Energy},
  volume = {217},
  number = {2},
  pages = {185--192},
  doi = {10.1243/09576500360611218},
  urldate = {2026-08-11},
  langid = {english}
}

@article{goswamiLearningStiffChemical2024a,
  title = {Learning Stiff Chemical Kinetics Using Extended Deep Neural Operators},
  author = {Goswami, Somdatta and Jagtap, Ameya D. and Babaee, Hessam and Susi, Bryan T. and Karniadakis, George Em},
  year = 2024,
  month = feb,
  journal = {Comput. Meth. Appl. Mech. Eng.},
  volume = {419},
  pages = {116674},
  issn = {0045-7825},
  doi = {10.1016/j.cma.2023.116674},
  langid = {english}
}

@article{gourianovQuantuminspiredApproachExploit2022,
  title = {A Quantum-Inspired Approach to Exploit Turbulence Structures},
  author = {Gourianov, Nikita and Lubasch, Michael and Dolgov, Sergey and van den Berg, Quincy Y. and Babaee, Hessam and Givi, Peyman and Kiffner, Martin and Jaksch, Dieter},
  year = 2022,
  month = jan,
  journal = {Nat. Comput. Sci.},
  volume = {2},
  number = {1},
  pages = {30--37},
  issn = {2662-8457},
  doi = {10.1038/s43588-021-00181-1},
  urldate = {2026-09-06},
  langid = {english}
}

@article{hanImprovedApproachMore2022,
  title = {An Improved Approach towards More Robust Deep Learning Models for Chemical Kinetics},
  author = {Han, Xu and Jia, Ming and Chang, Yachao and Li, Yaopeng},
  year = 2022,
  month = apr,
  journal = {Combust. Flame},
  volume = {238},
  pages = {111934},
  publisher = {Elsevier BV},
  issn = {0010-2180},
  doi = {10.1016/j.combustflame.2021.111934},
  urldate = {2025-07-25},
  copyright = {https://www.elsevier.com/tdm/userlicense/1.0/},
  langid = {english}
}

@article{ihmeCombustionMachineLearning2022,
  title = {Combustion Machine Learning: Principles, Progress and Prospects},
  shorttitle = {Combustion Machine Learning},
  author = {Ihme, Matthias and Chung, Wai Tong and Mishra, Aashwin Ananda},
  year = 2022,
  month = jul,
  journal = {Prog. Energy Combust. Sci.},
  volume = {91},
  pages = {101010},
  issn = {0360-1285},
  doi = {10.1016/j.pecs.2022.101010},
  langid = {english}
}

@article{jiangInterpretableFrameworkDatadriven2021,
  title = {An Interpretable Framework of Data-Driven Turbulence Modeling Using Deep Neural Networks},
  author = {Jiang, Chao and Vinuesa, Ricardo and Chen, Ruilin and Mi, Junyi and Laima, Shujin and Li, Hui},
  year = 2021,
  month = may,
  journal = {Phys. Fluids},
  volume = {33},
  number = {5},
  pages = {055133},
  issn = {1070-6631},
  doi = {10.1063/5.0048909},
  urldate = {2026-08-28},
  langid = {english}
}

@article{jiStiffPINNPhysicsinformedNeural2021,
  title = {Stiff-{{PINN}}: Physics-Informed Neural Network for Stiff Chemical Kinetics},
  shorttitle = {Stiff-{{PINN}}},
  author = {Ji, Weiqi and Qiu, Weilun and Shi, Zhiyu and Pan, Shaowu and Deng, Sili},
  year = 2021,
  month = aug,
  journal = {J. Phys. Chem. A},
  volume = {125},
  number = {36},
  pages = {8098--8106},
  issn = {1089-5639},
  doi = {10.1021/acs.jpca.1c05102},
  langid = {english}
}

@misc{kazakovDRM191995,
  author = {Kazakov, A. and Frenklach, M.},
  url = {http://www.me.berkeley.edu/drm/},
  urldate = {2026-09-02},
  langid = {english},
  lastchecked = {2 September 2026}
}

@article{kumarCombustionChemistryAcceleration2024,
  title = {Combustion Chemistry Acceleration with {{DeepONets}}},
  author = {Kumar, Anuj and Echekki, Tarek},
  year = 2024,
  month = jun,
  journal = {Fuel},
  volume = {365},
  pages = {131212},
  issn = {00162361},
  doi = {10.1016/j.fuel.2024.131212},
  urldate = {2025-07-28},
  langid = {english}
}

@article{liComprehensiveDeepLearning2025,
  title = {Comprehensive Deep Learning for Combustion Chemistry Integration: {{Multi-fuel}} Generalization and {\emph{a Posteriori}} Validation in Reacting Flow},
  shorttitle = {Comprehensive Deep Learning for Combustion Chemistry Integration},
  author = {Li, Han and Yang, Ruixin and Xu, Yangchen and Zhang, Min and Mao, Runze and Chen, Zhi X.},
  year = 2025,
  month = jan,
  journal = {Phys. Fluids},
  volume = {37},
  number = {1},
  pages = {015162},
  issn = {1070-6631, 1089-7666},
  doi = {10.1063/5.0248582},
  urldate = {2025-08-08},
  langid = {english}
}

@article{luAccommodatingRealisticFuel2009,
  title = {Toward Accommodating Realistic Fuel Chemistry in Large-Scale Computations},
  author = {Lu, Tianfeng and Law, Chung K.},
  year = 2009,
  month = apr,
  journal = {Prog. Energy Combust. Sci.},
  volume = {35},
  number = {2},
  pages = {192--215},
  issn = {0360-1285},
  doi = {10.1016/j.pecs.2008.10.002},
  langid = {english}
}

@article{murakamiC3MechLiteIntegratedComponent2025,
  title = {{{C3MechLite}}: An Integrated Component Library of Compact Kinetic Mechanisms for Low-Carbon, Carbon Neutral and Zero-Carbon Fuels},
  shorttitle = {{{C3MechLite}}},
  author = {Murakami, Yuki and Wang, Quan-De and Liu, Shuaishuai and Zhu, Yuxiang and Wang, Pengzhi and Maffei, Luna Pratali and Langer, Raymond and Faravelli, Tiziano and Pitsch, Heinz and Klippenstein, Stephen J and Bergthorson, Jeff and Bourque, Gilles and Wagnon, Scott and Senecal, Peter Kelly and Curran, Henry},
  year = 2025,
  month = dec,
  journal = {Combust. Flame},
  volume = {282},
  pages = {114410},
  issn = {0010-2180},
  doi = {10.1016/j.combustflame.2025.114410},
  langid = {english}
}

@article{NEURIPS2019_0234c510,
  title   = {Comparing Images Using the {Hausdorff} Distance},
  author  = {Huttenlocher, Daniel P. and Klanderman, Gregory A. and Rucklidge, William J.},
  journal = {IEEE Trans. Pattern Anal. Mach. Intell.},
  volume  = {15},
  number  = {9},
  year    = {1993},
  pages   = {850--863},
  doi     = {10.1109/34.232073},
}

@article{ramanEmergingTrendsNumerical2019,
  title = {Emerging Trends in Numerical Simulations of Combustion Systems},
  author = {Raman, Venkat and Hassanaly, Malik},
  year = 2019,
  month = jan,
  journal = {Proc. Combust. Inst.},
  volume = {37},
  number = {2},
  pages = {2073--2089},
  issn = {1540-7489},
  doi = {10.1016/j.proci.2018.07.121},
  langid = {english}
}

@article{readshawModelingTurbulentFlames2021,
  title = {Modeling of Turbulent Flames with the Large Eddy Simulation--Probability Density Function ({{LES}}--{{PDF}}) Approach, Stochastic Fields, and Artificial Neural Networks},
  author = {Readshaw, Thomas and Ding, Tianjie and Rigopoulos, Stelios and Jones, W. P.},
  year = 2021,
  month = mar,
  journal = {Phys. Fluids},
  volume = {33},
  number = {3},
  pages = {035154},
  issn = {1070-6631},
  doi = {10.1063/5.0041122},
  urldate = {2026-08-28},
  langid = {english}
}

@article{saitoDatadrivenSimulationAmmonia2023,
  title = {Data-Driven Simulation of Ammonia Combustion Using Neural Ordinary Differential Equations ({{NODE}})},
  author = {Saito, Manabu and Xing, Jiangkuan and Nagao, Jun and Kurose, Ryoichi},
  year = 2023,
  month = dec,
  journal = {Appl. Energy Combust. Sci.},
  volume = {16},
  pages = {100196},
  issn = {2666-352X},
  doi = {10.1016/j.jaecs.2023.100196},
  langid = {english}
}

@article{schultheisQuantitativeMeasurementsThermochemical2024,
  title = {Quantitative Measurements of Thermo-Chemical States in Turbulent Lean and Rich Premixed {{\ce{NH3}}}/{{\ce{H2}}}/{{\ce{N2}}}-air Jet Flames},
  author = {Schultheis, Robin and Li, Tao and Shi, Shuguo and Barlow, Robert S. and Zhou, Bo and Geyer, Dirk and Dreizler, Andreas},
  year = 2024,
  month = jan,
  journal = {Proc. Combust. Inst.},
  volume = {40},
  number = {1},
  pages = {105571},
  issn = {1540-7489},
  doi = {10.1016/j.proci.2024.105571},
  langid = {english}
}

@misc{smithGRIMech301999,
  author = {Smith, Gregory P. and Golden, David M. and Frenklach, Michael and Moriarty, Nigel W. and Eiteneer, Boris and Goldenberg, Mikhail and Bowman, C. Thomas and Hanson, Ronald K. and Song, Soonho and Gardiner, Jr., William C. and Lissianski, Vitali V. and Qin, Zhiwei},
  url = {http://www.me.berkeley.edu/gri_mech/},
  urldate = {2026-09-02},
  langid = {english},
  lastchecked = {2 September 2026}
}

@article{trefethenHydrodynamicStabilityEigenvalues1993,
  title = {Hydrodynamic Stability without Eigenvalues},
  author = {Trefethen, Lloyd N. and Trefethen, Anne E. and Reddy, Satish C. and Driscoll, Tobin A.},
  year = 1993,
  month = jul,
  journal = {Science},
  volume = {261},
  number = {5121},
  pages = {578--584},
  publisher = {American Association for the Advancement of Science},
  doi = {10.1126/science.261.5121.578},
  urldate = {2026-09-03},
  langid = {english}
}

@article{villaizan-valleladoDiffusionModelsTabular2025,
  title = {Diffusion Models for Tabular Data Imputation and Synthetic Data Generation},
  author = {{Villaiz{\'a}n-Vallelado}, Mario and Salvatori, Matteo and Segura, Carlos and Arapakis, Ioannis},
  year = 2025,
  month = jul,
  journal = {ACM Trans. Knowl. Discov. Data},
  volume = {19},
  number = {6},
  pages = {1--32},
  issn = {1556-4681, 1556-472X},
  doi = {10.1145/3742435},
  urldate = {2026-08-16},
  langid = {english}
}

@article{wengExtendedFourierNeural2025,
  title = {Extended {{Fourier Neural Operators}} to Learn Stiff Chemical Kinetics under Unseen Conditions},
  author = {Weng, Yuting and Li, Han and Zhang, Hao and Chen, Zhi X. and Zhou, Dezhi},
  year = 2025,
  month = feb,
  journal = {Combust. Flame},
  volume = {272},
  pages = {113847},
  issn = {00102180},
  doi = {10.1016/j.combustflame.2024.113847},
  urldate = {2026-05-06},
  langid = {english}
}

@article{yangDiffusionModelsComprehensive2024,
  title = {Diffusion Models: A Comprehensive Survey of Methods and Applications},
  shorttitle = {Diffusion Models},
  author = {Yang, Ling and Zhang, Zhilong and Song, Yang and Hong, Shenda and Xu, Runsheng and Zhao, Yue and Zhang, Wentao and Cui, Bin and Yang, Ming-Hsuan},
  year = 2024,
  month = apr,
  journal = {ACM Comput. Surv.},
  volume = {56},
  number = {4},
  pages = {1--39},
  issn = {0360-0300, 1557-7341},
  doi = {10.1145/3626235},
  urldate = {2026-08-16},
  langid = {english}
}

@article{zhangCRKPINNPhysicsinformedNeural2024,
  title = {{{CRK-PINN}}: A Physics-Informed Neural Network for Solving Combustion Reaction Kinetics Ordinary Differential Equations},
  shorttitle = {Crk-Pinn},
  author = {Zhang, Shihong and Zhang, Chi and Wang, Bosen},
  year = 2024,
  month = nov,
  journal = {Combust. Flame},
  volume = {269},
  pages = {113647},
  issn = {0010-2180},
  doi = {10.1016/j.combustflame.2024.113647},
  langid = {english}
}

@article{zhangLearningTransientEvolution2024,
  title = {Learning Transient Evolution of Multidimensional Reacting Flows by Multiscale {{Fourier}} Neural Operators},
  author = {Zhang, Hao and Weng, Yuting and Zhao, Zhiwei and Zhou, Dezhi},
  year = 2024,
  journal = {Proc. Combust. Inst.},
  volume = {40},
  number = {1-4},
  pages = {105714},
  issn = {15407489},
  doi = {10.1016/j.proci.2024.105714},
  urldate = {2026-07-16},
  langid = {english}
}

@article{zhangMultiscaleSamplingMethod2022,
  title = {A Multi-Scale Sampling Method for Accurate and Robust Deep Neural Network to Predict Combustion Chemical Kinetics},
  author = {Zhang, Tianhan and Yi, Yuxiao and Xu, Yifan and Chen, Zhi X. and Zhang, Yaoyu and E, Weinan and Xu, Zhi-Qin John},
  year = 2022,
  month = nov,
  journal = {Combust. Flame},
  volume = {245},
  pages = {112319},
  issn = {0010-2180},
  doi = {10.1016/j.combustflame.2022.112319},
  langid = {english}
}

@article{zhengRFNOPINNPhysicsinformedNeural2026,
  title = {{{RFNO-PINN}}: A Physics-Informed Neural Operator for Predicting Multi-Species Combustion Dynamics},
  shorttitle = {Rfno-Pinn},
  author = {Zheng, Yang and Xu, Leilei and Wu, Xieming and Zhou, Yuchen and Zhou, Dezhi and Lu, Xingcai and Bai, Xue-Song},
  year = 2026,
  journal = {Proc. Combust. Inst.},
  volume = {42},
  pages = {106152},
  issn = {15407489},
  doi = {10.1016/j.proci.2026.106152},
  urldate = {2026-07-13},
  langid = {english}
}

@article{zhangPerformanceComparisonMachine2026,
	title = {Performance comparison of machine learning and {GPU} accelerated solvers in high-fidelity simulation of {Cambridge} stratified flames},
	volume = {426},
	issn = {0016-2361},
	doi = {10.1016/j.fuel.2026.139562},
	journal = {Fuel},
	author = {Zhang, Min and Dang, Guanlin and Mao, Runze and Cai, Yuqing and Yang, Ruixing and Li, Han and Chen, Zhi X.},
	month = dec,
	year = {2026},
	pages = {139562},
}

\onecolumn
\begingroup
\newgeometry{twocolumn=false,left=90bp,right=90bp,top=72bp,textheight=697.89bp,headheight=0pt,headsep=0pt,footskip=0pt}
\setlength{\paperheight}{297mm}
\pdfpageheight=\paperheight
\pagestyle{empty}
\raggedbottom
\expandafter\def\csname u8:：\endcsname{\makebox[1em][c]{:}}
\fontfamily{ptm}\selectfont
\fontsize{11bp}{18.2bp}\selectfont
\setlength{\parindent}{0pt}
\setlength{\parskip}{8bp}
\setlength{\emergencystretch}{1em}
\hyphenpenalty=10000
\exhyphenpenalty=50

{\fontsize{17bp}{20bp}\selectfont \textbf{Supplementary Material}\par}

{\fontsize{13bp}{15bp}\selectfont Dong Yang\textsuperscript{a}, Xiongbo Duan\textsuperscript{a}, Min Zhang\textsuperscript{a,}\textsuperscript{\ensuremath{\ast}}, Yuqiang Li\textsuperscript{a}\par}

{\fontsize{8bp}{10bp}\selectfont \textit{a }\textit{School of Energy Science and Engineering, Central South University, Changsha, 410083, People’s Republic of China }\par}

\textbf{Supplementary A：}Ablation of thermochemical loss terms\par

Figure SA. 1 compares the complete TC-MLNO model with variants omitting \textit{L}\textsubscript{Mass}, \textit{L}\textsubscript{Elements}, or \textit{L}\textsubscript{HRR} for one-dimensional flames simulated using GRI-Mech 3.0. All four cases reproduce the overall flame structure, while the peak-region insets reveal differences in species and HRR predictions. The complete model yields the smallest errors in the NNH and HRR peak magnitudes at both equivalence ratios. The NNH peak errors are 0.27\% and 0.24\% at \textit{\ensuremath{\phi}} = 0.95 and 1.05, respectively, while the corresponding HRR peak errors are 0.21\% and 0.19\%. Removing \textit{L}\textsubscript{Mass} produces the largest profile deviations and shifts the peaks toward the unburned-gas region. Specifically, removing \textit{L}\textsubscript{Mass} increases the relative \textit{L}\textsubscript{2} error of HRR from 2.83\% to 7.61\% at \textit{\ensuremath{\phi}} = 0.95 and from 0.21\% to 6.07\% at \textit{\ensuremath{\phi}} = 1.05. The equal-weight mean of these relative \textit{L}\textsubscript{2} errors across the profiles is 1.07\% for the complete model, compared with 1.21\%, 2.08\%, and 4.68\% for the variants without \textit{L}\textsubscript{HRR}, \textit{L}\textsubscript{Elements}, and \textit{L}\textsubscript{Mass}, respectively. The complete model therefore demonstrates better predictive performance for the flame structure.\par

\noindent\includegraphics[width=396.85039bp]{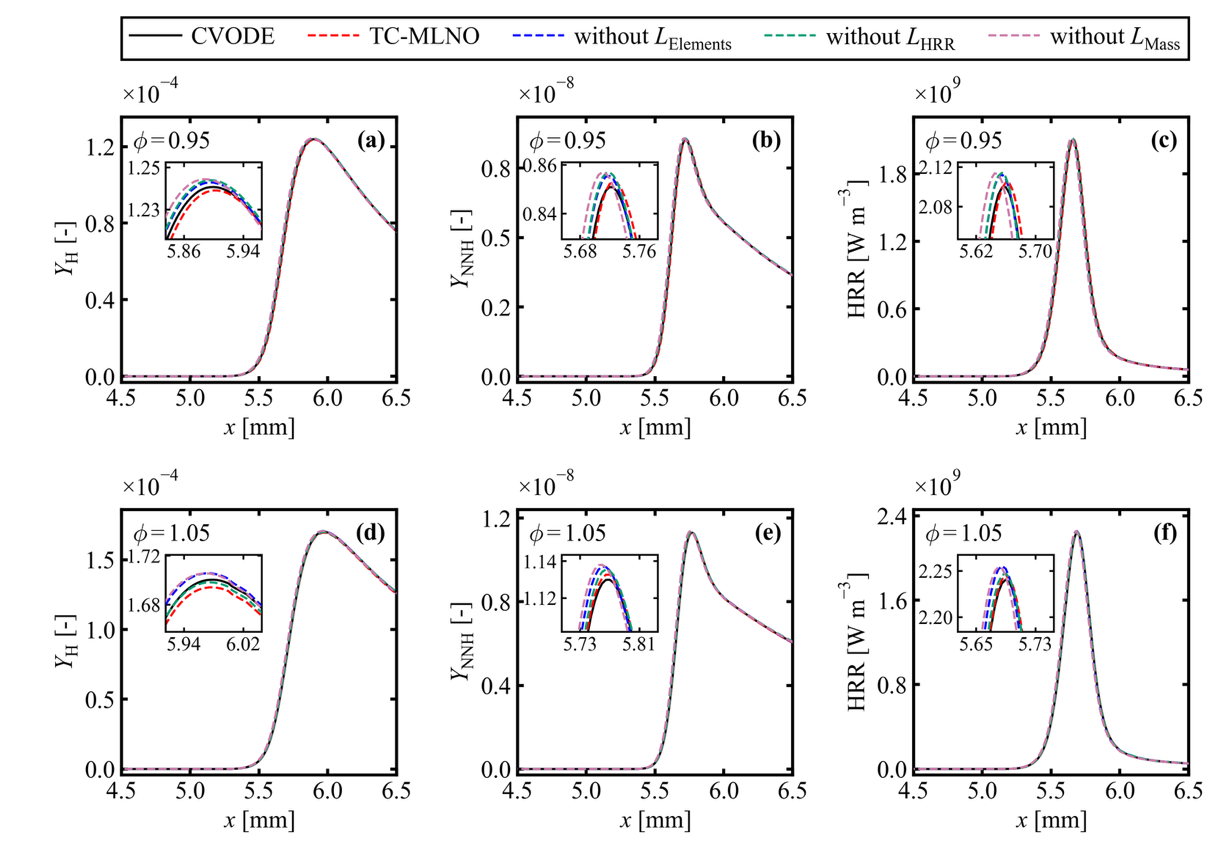}\par

{\fontsize{10bp}{18.2bp}\selectfont Fig SA. 1 Loss-term ablation of H, NNH, and heat release rate (HRR) profiles at \textit{\ensuremath{\phi}}\textit{ }= 0.95 and 1.05.\par}

\newpage

\textbf{Supplementary B：}\textbf{ }Species-resolved errors for random and diffusion sampling\par

Figure SB. 1 further resolves the relative\textit{ L}\textsubscript{\textit{2}} errors into temperature, major species, and radicals. The diffusion-trained model has a lower error for every displayed quantity, confirming that the reduction is not caused by only a few species. Its temperature error is 0.1\%, while the errors of CH\textsubscript{4}, NH\textsubscript{3}, and O\textsubscript{2} are approximately 0.03\%. More pronounced differences occur for quantities with peaks or extended post-flame tails when adopting the diffusion model sampling: the CO error decreases from 13.9\% to 0.8\%, the OH error from 9.4\% to 1.9\%, and the HCO error from 6.7\% to 0.3\%.\par

\vspace{12.08bp}
\noindent\includegraphics[width=376.00000bp]{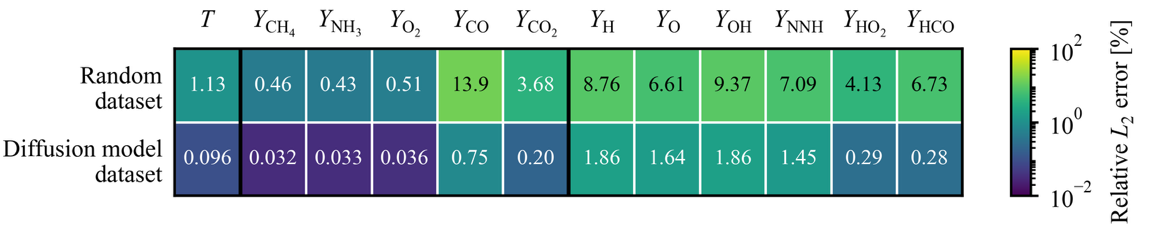}\par

{\fontsize{10bp}{18.2bp}\selectfont Fig SB. 1 Relative $L_2$ errors of temperature and representative species for the GRI-Mech 3.0 one-dimensional flame.\par}

\vspace{25.4bp}

\textbf{Supplementary C：} Reaction-timescale grouping of GRI-Mech 3.0 species\par

The 52 predicted species are grouped using their characteristic destruction times. For each species, the median of log10(\textit{\ensuremath{\tau}}\textsubscript{\textit{k}}/\ensuremath{\Delta}\textit{t}) over representative training states is used for ordering, where \ensuremath{\Delta}\textit{t} is the network time increment. To balance group sizes and timescale separation, gap-based grouping is used when the maximum adjacent gap is at least 0.5 orders of magnitude, with boundaries chosen to maximize their combined gaps under balanced-size constraints. The median values range from \ensuremath{-}1.98 for NNH to 6.80 for N\textsubscript{2}, spanning 8.78 orders of magnitude relative to the network step. The fast group contains 16 species from NNH to C\textsubscript{3}H\textsubscript{7} over \ensuremath{-}1.98–0.45. The medium group contains 16 species from OH to HCCOH over 0.56–2.91. The slow group contains 20 species from CH\textsubscript{3}OH to N\textsubscript{2} over 3.23–6.80. Each group is represented by a separate TC-MLNO branch.\par

\newpage

\noindent\includegraphics[width=410.40000bp]{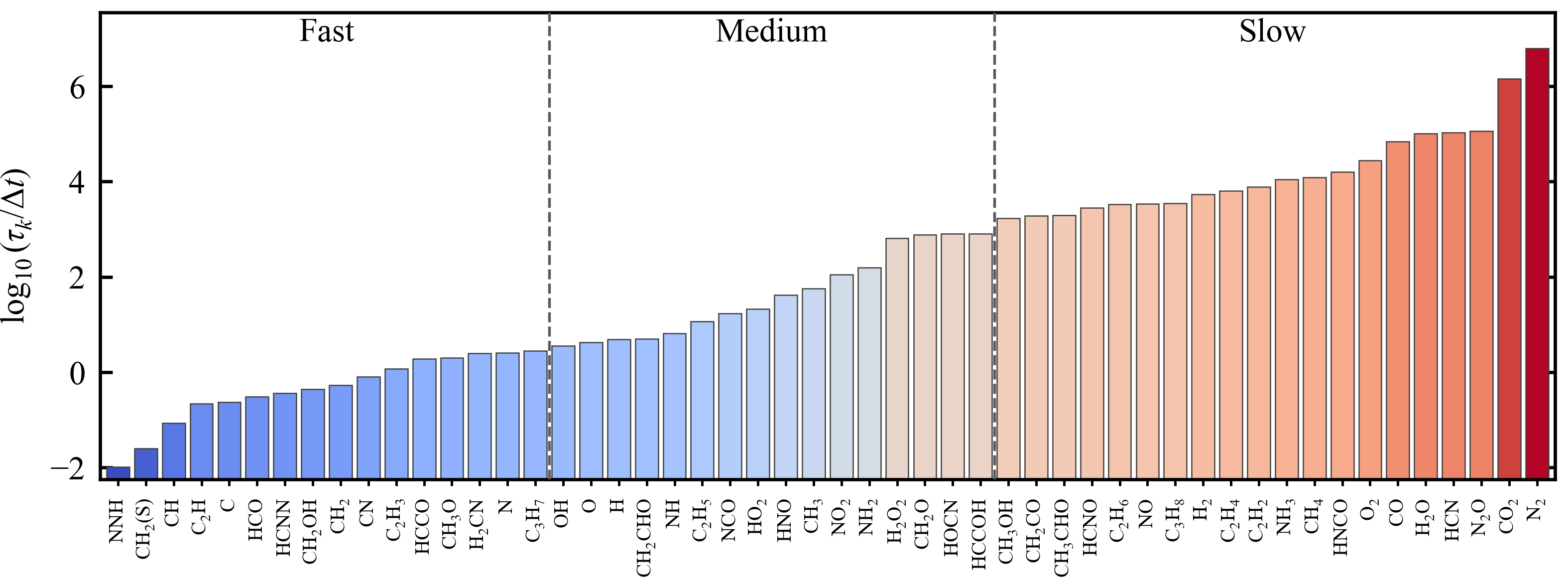}\par
\vspace{-5.55bp}

{\fontsize{10bp}{18.2bp}\selectfont Fig SC. 1 Reaction-time-scale ordering of the 52 predicted GRI-Mech 3.0 species. Dashed vertical lines indicate the fast, medium,and slow groups containing 16, 16, and 20 species, respectively.\par}

\clearpage
\endgroup

\end{document}